\documentclass[aip,rsi,reprint,graphicx]{revtex4-1} 

\usepackage{amsmath}    
\usepackage{verbatim}   
\usepackage{color}      
\usepackage{subfigure}  
\usepackage{placeins}	
\usepackage[pdftex]{graphicx} 
\usepackage{epstopdf}
\usepackage{hyperref}
\hypersetup{
  colorlinks=true,
  linkcolor=blue,
  citecolor=blue,
  urlcolor=blue,
  breaklinks=true}

\begin{document}

\title{A highly stable monolithic enhancement cavity for SHG generation in the UV} 

\author{S. Hannig}
\affiliation{Physikalisch-Technische Bundesanstalt, Bundesallee 100, 38116 Braunschweig, Germany}

\author{J. Mielke}
\affiliation{Institut f{\"u}r Quantenoptik, Leibniz Universit{\"a}t Hannover, Welfengarten 1, 30167 Hannover, Germany}

\author{J. A. Fenske}
\affiliation{Physikalisch-Technische Bundesanstalt, Bundesallee 100, 38116 Braunschweig, Germany}

\author{M. Misera}
\affiliation{Physikalisch-Technische Bundesanstalt, Bundesallee 100, 38116 Braunschweig, Germany}

\author{N. Beev}
\altaffiliation{current address: CERN}
\affiliation{Physikalisch-Technische Bundesanstalt, Bundesallee 100, 38116 Braunschweig, Germany}

\author{C. Ospelkaus}
\affiliation{Physikalisch-Technische Bundesanstalt, Bundesallee 100, 38116 Braunschweig, Germany}
\affiliation{Institut f{\"u}r Quantenoptik, Leibniz Universit{\"a}t Hannover, Welfengarten 1, 30167 Hannover, Germany}

\author{P. O. Schmidt}
\email[Corresponding author: ]{Piet.Schmidt@ptb.de}
\affiliation{Physikalisch-Technische Bundesanstalt, Bundesallee 100, 38116 Braunschweig, Germany}
\affiliation{Institut f{\"u}r Quantenoptik, Leibniz Universit{\"a}t Hannover, Welfengarten 1, 30167 Hannover, Germany}

\date{\today}

\begin{abstract}
We present a highly stable bow-tie power enhancement cavity for critical second-harmonic generation into the UV using a Brewster-cut $\beta$-BaB$_2$O$_4$ (BBO) nonlinear crystal. The cavity geometry is suitable for all UV wavelengths reachable with BBO and can be modified to accommodate anti-reflection coated crystals, extending its applicability to the entire wavelength range accessible with non-linear frequency conversion. The cavity is length-stabilized using a fast general purpose digital PI controller based on the open source STEMlab 125-14 (formerly Red Pitaya) system acting on a mirror mounted on a fast piezo actuator. We observe $130\,\mathrm{h}$ uninterrupted operation without decay in output power at $313\,\mathrm{nm}$. The robustness of the system has been confirmed by exposing it to accelerations of up to $1\,\mathrm{g}$ with less than $10\%$ in-lock output power variations. Furthermore, the cavity can withstand 30~minutes of acceleration exposure at a level of $3\,\mathrm{g}_\mathrm{rms}$ without substantial change in SHG output power, demonstrating that the design is suitable for transportable setups.
\end{abstract}

\pacs{42.60.Da, 42.60.Lh, 42.60.Pk, 42.62.Eh, 42.65.Ky, 42.79.Nv}

\maketitle

\section{Introduction}
\label{sec:intro}
Today's quantum optics experiments require a broad range of laser frequencies. Usually, several continuous wave laser systems are required per experiment \cite{tan_multi-element_2015,chou_frequency_2010,yamanaka_frequency_2015} which imposes tight bounds on the reliability and cost per system. Recent progress in the development of quantum sensors has led to systems outperforming classical devices, such as quantum gravimeters \cite{hauth_first_2013, barrett_mobile_2013}, and clocks exceeding $10^{-17}$ accuracy \cite{huntemann_single-ion_2016, chou_frequency_2010, nicholson_systematic_2015}. New applications such as relativistic geodesy \cite{bjerhammar_relativistic_1985} using transportable optical clocks \cite{koller_transportable_2017, grotti_geodesy_2017, cao_transportable_2016} or taking advantage of the long interaction times in atom interferometry experiments in a microgravity or even space environment \cite{muntinga_interferometry_2013, origlia_development_2016, yin_satellite-based_2017} have emerged. For these applications, quantum optics experiments need to be operated outside highly-specialized laboratories, increasing the demands in terms of mechanical robustness of the optical setups.
Small and at the same time reliable laser sources are available only for a restricted wavelength range \cite{luvsandamdin_micro-integrated_2014,kohfeldt_compact_2016}. To reach other wavelengths, non-linear conversion processes, such as sum- or difference frequency or harmonic generation \cite{wakui_generation_2014, hu_high_2013, sherstov_diode-laser_2010, scheid_750_2007, wen_cavity-enhanced_2014, eismann_all-solid-state_2012, wilson_750-mw_2011, vasilyev_compact_2011, carollo_third-harmonic-generation_2017} are typically employed. A common approach to generate the desired wavelength is second harmonic generation \cite{Franken_Optical_Harmonics_1961} (SHG) of External Cavity Diode Lasers (ECDLs) or fiber lasers. The frequency doubling of infrared (IR) lasers has been demonstrated in single-crystal monolithic ring cavities \cite{kane_monolithic_1985,kozlovsky_efficient_1988,gerstenberger_efficient_1991}. With the advent of commercially available periodically-poled wave\-guide doublers, e.g. based on Lithium Tantalate or Lithium Niobate nonlinear crystals, wavelength conversion into the blue spectral range has been achieved in compact setups \cite{NTT}. Monolithic ring cavities have been demonstrated to produce blue light down to wavelengths of $429\,\mathrm{nm}$ \cite{kozlovsky_blue_1994,hemmerich_compact_1994,skoczowsky_efficient_2010}.
However, to our knowledge neither single-crystal monolithic ring cavities nor modules are available for UV generation below 350~nm. For these UV wavelengths SHG in nonlinear crystals such as Beta-Barium-Borate ($\beta-\mathrm{BaB}_2\mathrm{O}_4$, BBO), placed in an optical enhancement resonator, are typically employed\cite{ZI1990}. Commercially available systems usually come with restricted flexibility, e.g. lacking access to the SHG light internally reflected at the output facet of Brewster-cut nonlinear crystals. In contrast, self-built systems made from off-the-shelve components usually lack mechanical stability and re\-li\-a\-bil\-i\-ty.

Traditionally, the length of the cavity is kept interferometrically stable by displacing one of the cavity mirrors in a proportional-integral (PI) feedback loop usually implemented using analog electronics. With the availability of fast general-purpose digital hardware such as Field Programmable Gate Arrays (FPGAs) and/or microcontroller-based systems it has become popular to employ digital feedback controllers instead \cite{HU2014,DB2009,sparkes_scalable_2011,LH2015,DL110}. These devices are more universally applicable, feature higher usability, and can easily be linked to existing experimental control infrastructure.

Here we report on a mechanical monolithic bow-tie cavity design \cite{AS2005} for critically phase-matched SHG generation in a BBO crystal and its general purpose digital PI locking electronics based on a modified STEMlab 125-14 application \cite{RedPit,JF15}. The cavity is implemented in a robust monolithic mechanical support frame and equipped with a minimal set of high-quality adjustment screws that are accessible from outside. The cavity geometry is suitable for the generation of all wavelengths accessible with BBO in SHG ooe-processes, starting at $204.8\,\mathrm{nm}$ \cite{kato_second-harmonic_1986}. The internal reflections of the pump light (PL) at the fundamental frequency and SHG on the crystal's facets are accessible through two additional windows. In order to prevent moisture-induced fogging of the crystal, a dry purging gas, such as nitrogen or oxygen can be applied inside the sealed cavity. 

We measure a locking bandwidth of $17\,\mathrm{kHz}$ and demonstrate $130\,\mathrm{h}$ continuous operation in lock for the conversion from $626\,\mathrm{nm}$ to $313\,\mathrm{nm}$ without substantial decay in the output power. Moreover, the cavity remains in lock while being exposed to accelerations in the vertical direction of around $1\,\mathrm{g}$. A comparison of the output power before and after a $30\,\mathrm{min}$ high-acceleration shaking in one dimension demonstrates that the cavity can withstand a truck transport according to ISO13355:2016 without significant change in alignment. 

The paper is structured as follows: Section~\ref{sec:theory} briefly summarizes the theory of SHG and bow-tie power enhancement cavities. In Section~\ref{sec:cavdesign} we describe the monolithic cavity design and in Section~\ref{sec:setup} the setup of the complete system including the STEMlab-based locking electronics is described. The results on acceleration testing and long term stability are presented in Section~\ref{sec:experiment}.

\section{Theoretical Background}
\label{sec:theory}
When light passes through a nonlinear medium, SHG light of twice the frequency can be generated. The nonlinear conversion process scales with the square of the PL intensity. In the case of low conversion efficiency and therefore undepleted PL, the SHG power $P_\mathrm{SHG}$ scales with the square of the pump light power $P_\mathrm{PL}$: 
\begin{align}
\label{eq:shg}
P_\mathrm{SHG}=\kappa P_\mathrm{PL}^2,
\end{align}
where $\kappa$ is the conversion coefficient. Efficient SHG generation is only achieved if the wavevectors of the pump light $\vec{k}_\mathrm{PL}$ and the SHG $\vec{k}_\mathrm{SHG}$ fulfill the phasematching condition
\begin{align}
\vec{k}_\mathrm{SHG}=2\vec{k}_\mathrm{PL}.
\end{align}
In BBO this can be achieved by taking advantage of the crystal's birefringence with different refractive indices $n_o$ and $n_e$ for different incident polarisations. By choosing the appropriate phase-matching angle $\theta_\mathrm{pm}$ within the resulting index ellipsoid, $n_\mathrm{SHG}=n_\mathrm{PL}$ can be achieved \cite{nikogosyan_nonlinear_2005}. For SHG of $626\,\mathrm{nm}$ in BBO the resulting phase-matching angle is 
\begin{align}
\theta_\mathrm{pm}=38.4^\circ.
\end{align}

\subsection{SHG power optimization}
\label{subsec:script}
The single-pass conversion efficiency of cw light in a BBO crystal of a few mm length general is rather low (on the order of $\kappa\sim 10^{-4}$\,1/W). Therefore, an enhancement cavity for the PL is built around the nonlinear crystal \cite{Franken_Optical_Harmonics_1961, armstrong_interactions_1962, Ashkin_Resonant_SHG_1966}. The optimization of the cavity SHG output power can be divided into two steps: first the optimization of the power generated per single pass of the PL through the crystal\cite{boyd_parametric_1968} and second the optimization of the power enhancement \cite{jurdik_performance_2002}.

\subsubsection{Single pass optimization}
\label{subsubsec:spopt}
As mentioned above, the SHG power generated per infinitesimal crystal volume is proportional to the square of the PL intensity. Tighter focusing of the PL enhances the intensity at the focus, at the expense of lower intensity away from the focus. This tradeoff for a given crystal length $l$ has been investigated for the propagation of a circular Gaussian beam with minimum waist $w_0$ by Boyd and Kleinman\cite{boyd_parametric_1968}, who derived an optimum focusing ratio of $l/b=2.84$ in the absence of birefringence, where $l$ is the crystal length and $b=w_0^2k$ the confocal parameter with $k=2\pi/\lambda$. However, in the presence of birefringence, the phase vector $\vec{k}$ and Poynting vector $\vec{S}$ of the extraordinary SHG wave are in general not parallel, which results in a walk-off angle $\varrho$ between SHG and PL beam of
\begin{align}
\label{eq:woa}
\varrho=4.6^\circ
\end{align}
for SHG of $626\,\mathrm{nm}$ in BBO. This effect is quantified by the walk-off parameter $B$\cite{boyd_parametric_1968}:
\begin{align}
\label{eq:awo}
B=\varrho\sqrt{lk}/2 
\end{align}
A strong walk-off (such as in BBO), requires weaker focusing of the pump light to optimize the spatial overlap between PL and SHG.
Furthermore, a focused Gaussian beam exhibits a phase deviation $\Delta k$ from an ideal plane wave that changes along the propagation direction, known as Gouy effect\cite{Gouy_1890}. It is typically quantified by the parameter $\sigma=\frac{1}{2}b\Delta k$. For the SHG considered here, this means that the degree to which the phasematching condition is fulfilled changes over the crystal length. Taking all these effects into account through integration over infinitesimal contributions of the PL to the SHG light when propagating through the crystal, the Boyd-Kleinman theory provides optimum parameters $l/b$ and $\sigma$ for single-pass SHG output power. 
For SHG from $626\,\mathrm{nm}$ to $313\,\mathrm{nm}$ in a $10\,\mathrm{mm}$ long BBO crystal, we numerically obtain
$l/b=1.42$, $\sigma=0.75$, and $\omega_0=19\,\mu\mathrm{m}$ with a single-pass conversion efficiency of $\kappa=1.1\cdot10^{-4}$. A simlar analysis can be performed in case of an elliptical focus inside the crystal \cite{Freegarde:97}, as is the case for astigmatic cavities and/or Brewster-cut crystals.

\subsubsection{Cavity geometry}
\label{subsubsec:csopt}
\begin{figure}
\includegraphics[width=8.0cm]{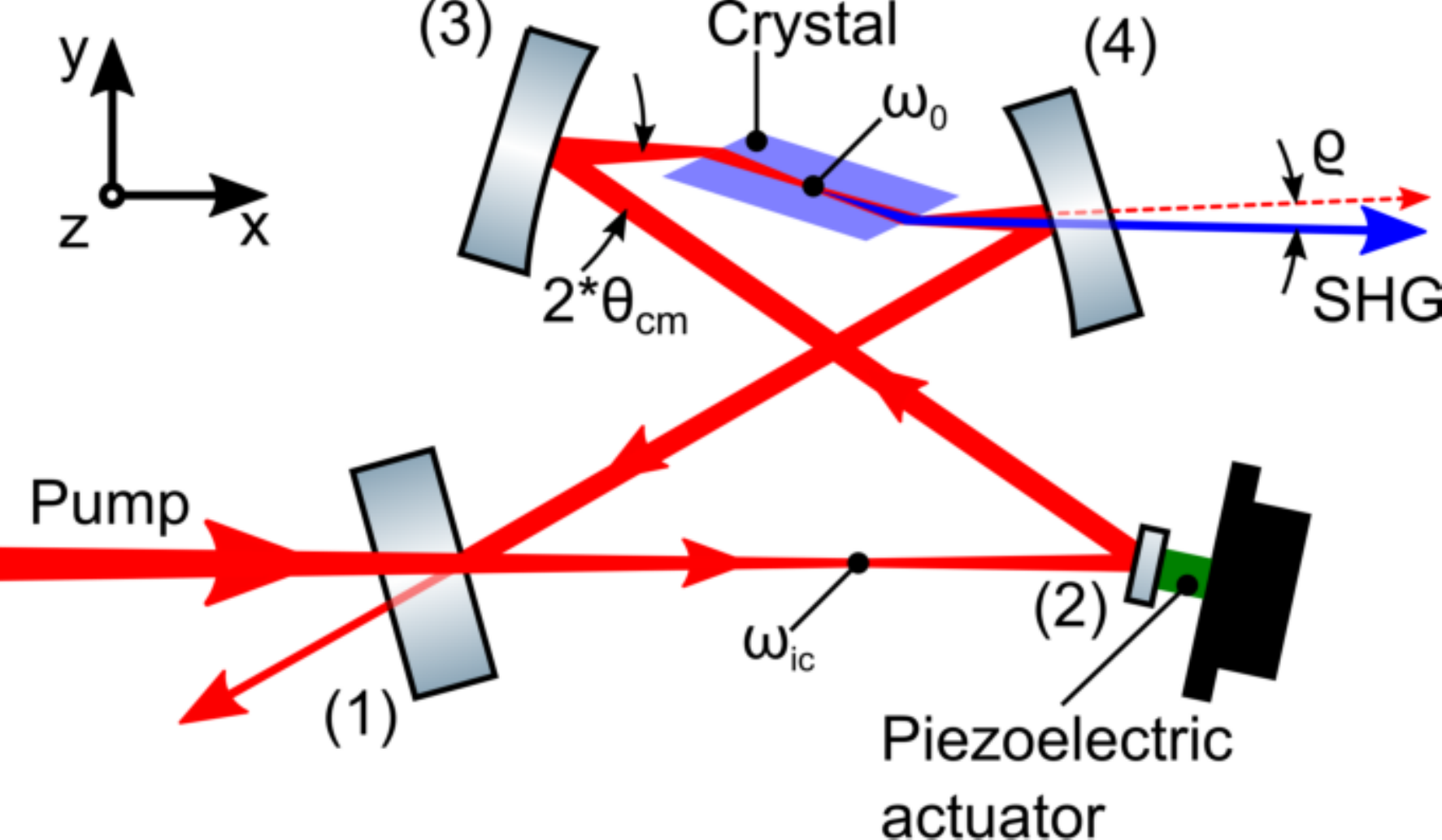}
\caption{Schematic layout of a bow-tie cavity. The resonator is pumped through mirror (1) with light focused on the incoupling waist $\omega_{ic}$. A lightweight mirror (2) on a piezoelectric actuator is used to lock the length of the cavity. (3) and (4) focus the light to the waist $\omega_0$ inside the nonlinear crystal and modematch the beam for the next roundtrip.}
\label{fig:bopri}
\end{figure}

A nonlinear medium with low conversion efficiency placed in an optical cavity in which the PL power is enhanced by a factor on the order of 50, can provide up to 2500 times more SHG output power than is possible in single pass configuration. 

A common design for an optical enhancement cavity, referred to as ``bow-tie'', consists of four mirrors and is schematically shown in Fig.~\ref{fig:bopri}. Compared to a linear cavity made of two mirrors, this geometry avoids the formation of a standing PL wave and thus reduces the likelyhood of photo-refractive effects. Furthermore, it allows for astigmatism compensation of Brewster-cut crystals by choosing a suitable distance between the two mirror pairs involved (see below). The incoupling mirror (1) fulfills the impedance matching condition (see below). Mirror (2) is mounted on a piezoelectric actuator with a maximum stroke sufficient to change the cavity length by more than one free spectral range (FSR) to enable length stabilization of the cavity to the PL wavelength. It is lightweight to achieve high mechanical resonance frequencies and thus a high feedback bandwidth. Mirrors (3) and (4) are concave, and their distance is chosen to provide a focus of the appropriate size at the center of the crystal. The fourth mirror also acts as outcoupler for the SHG light. A fully monolithic bow-tie cavity design similar to the Nd:YAG NPRO design\cite{kane_monolithic_1985} would suffer from detrimental PL loss due to the high linear absorption of $\alpha_{PL}\sim 0.01\,\mathrm{cm}^{-1}$ for $\lambda=626\,\mathrm{nm}$ in BBO \cite{Wata1991}.

\subsubsection{Crystal shape}
\label{subsubsec:cryshape}
For maximum PL power enhancement and therefore a high SHG output power (cf. Eq. \eqref{eq:shg}), low PL loss inside the cavity from absorption or scattering is required. 
One major source of loss in circulating PL is reflection at the interfaces of the nonlinear medium and the surrounding gas. These reflections can either be avoided by the application of an antireflective coating on the surfaces or by cutting and mounting the crystal under Brewster's angle to the incident beam. While AR coatings protect hygroscopic crystals like BBO against moisture in the environment, they are prone to damage induced by UV or pump light. Furthermore, residual reflections on surfaces orthogonal to the PL beam can cause a backward traveling PL wave. Weighting these properties and given the option to operate the crystal in a dry environment, it seems advantageous to choose a crystal cut under Brewster's angle $\theta_\mathrm{B}$ for the PL. For SHG of $626\,\mathrm{nm}$ in BBO the resulting angle is
\begin{align}
\theta_\mathrm{B}=\arctan(n_\mathrm{PL})=59.0^\circ,
\end{align}
where $n_\mathrm{PL}$ is the index of refraction for the PL.
Since PL and SHG have orthogonal polarization in ooe SHG, the Brewster condition can only be fulfilled for the pump light. The corresponding fractional internal reflection of the SHG at the crystal-gas-interface is given by Fresnel's equations for s-polarized light that simplify to
\begin{align}
R_\mathrm{SHG}=\left(\sin(\pi/2-2\theta_\mathrm{B})\right)^2\approx22\%
\end{align}

\subsubsection{Impedance matching}
\label{subsubsec:imat}
Assuming ideal mirrors and input beam mode matching, the entire pump light can be coupled into the cavity, if the incoupling mirror transmission is chosen in such a way that coupled pump light compensates all losses inside the cavity during a round trip.
In the case of a SHG cavity, the latter consist of two parts: PL loss due to conversion and parasitic loss due to imperfect reflection inside the resonator and linear absorption in the nonlinear crystal and surrounding gas with absorption coefficient $\alpha_\mathrm{PL}$.

The total round trip transmission of the cavity as a function of the single cavity mirror reflectivity $R_M$ and the crystal facet transmission $T_\mathrm{C}$ without considering conversion into SHG is
\begin{align}
\label{eq:rtt}
t=R_\mathrm{M}^3 T_\mathrm{C}^2(1-2l\alpha_\mathrm{PL})
\end{align}

Taking into account the loss through conversion with a factor $\kappa P_\mathrm{PL}^2$, the optimal incoupling mirror reflectivity $R_{in}$ for impedance matching is\cite{polzik_frequency_1991}
\begin{align}
\label{eq:rin}
R_\mathrm{in}=1-\left(\frac{1-t}{2}+\sqrt{\frac{(1-t)^2}{4}+t\kappa P_\mathrm{PL}}\right),
\end{align}
where the small correction factor $t$ from the incoupling mirror has been applied.

\subsubsection{Cavity parameters}
\label{subsubsec:cageo}
Both, the non-normal incidence of the cavity mode on the curved mirrors and crystal facets cut under Brewster's angle, introduce astigmatism in the cavity which needs to be accounted for in the design. The system forms two effective cavities in the $xy$ (sagittal) and $xz$ (tangential) planes which both need to be stable. Optimizing for simultaneous stability and conversion efficiency\cite{wilson_750-mw_2011,Freegarde:97} for a 10~mm BBO crystal and a 50~mm radius of curvature of the mirrors yields a mechanical cavity footprint which is inconveniently large to realize experimentally in a compact way. Instead, we choose the cavity parameters outlined in Tab. \ref{tab:626geo}. $w_{0,xy}$ and $w_{0,xz}$ are the saggital and tangential waist, respectively, which are larger than the optimum waist $\omega_0$ from Boyd-Kleinmann. This has the added advantage of reducing crystal damage from high intensities at the crystal surfaces. In the end the actual cavity geometry is defined by $l$, $\theta_\mathrm{B}$, the mirror to crystal distance $d_\mathrm{mc}$, $\theta_\mathrm{cm}$, and the geometric round-trip length $l_\mathrm{geo}$. 

\begin{table}[htbp!]
\caption{Geometric parameters of the doubling cavity using a $10\,\mathrm{mm}$ BBO-crystal pumped with $P_\mathrm{PL}=0.5\,\mathrm{W}$ at $\lambda_\mathrm{PL}=626\,\mathrm{nm}$ and $r=50\,\mathrm{mm}$ mirrors}
\label{tab:626geo}
\begin{small}
\begin{center}
\begin{ruledtabular}
\begin{tabular}{ll} 
Phasematching angle & $\theta_\mathrm{pm}=38.4^\circ$  
\tabularnewline
Waist saggital& $\omega_{0,xy}=25.3\,\mu\mathrm{m}$
\tabularnewline
Waist tangential& $\omega_{0,xz}=39.6\,\mu\mathrm{m}$  
\tabularnewline
Brewster angle & $\theta_\mathrm{B}=59.0^\circ$  
\tabularnewline
Angle on curved mirrors & $\theta_\mathrm{cm}=15.7^\circ$
\tabularnewline
Distance mirror to crystal & $d_\mathrm{mc}=25.1\,\mathrm{mm}$  
\tabularnewline
Geometric round trip length & $l_\mathrm{geo}=304.8\,\mathrm{mm}$  
\tabularnewline
Input coupling waist & $\omega_{ic}=166\,\mu\mathrm{m}$  
\tabularnewline

\end{tabular}
\end{ruledtabular}
\end{center}
\end{small}
\end{table}

For this choice of parameters, both the sagittal and tangential cavity are stable simultaneously and the input coupling waist (between the two flat mirrors) is nearly circular with a waist of $166\,\mu\mathrm{m}$. Compared to the geometry for optimal efficiency\cite{wilson_750-mw_2011,Freegarde:97}, the expected loss in conversion efficiency is only about $12\%$. This moderate loss is outweighed by the gain in mechanical stability and ruggedness due to the smaller footprint. The waists and Brewster's angle have been calculated for SHG of $\lambda=626\,\mathrm{nm}$, but their dependence on the PL wavelength is rather weak. For $\lambda=534\,\mathrm{nm}$ PL for instance, the waists change by $-1.9\,\mu\mathrm{m}$ and $-2.9\,\mu\mathrm{m}$. Brewster's angle increases by $0.1^\circ$. Therefore, the cavity geometry is practically suitable for frequency doubling of a spectral range of several hundred nm, as long as a  $10\,\mathrm{mm}$ long BBO-crystal cut under the corresponding phasematching angle is used. In conclusion, the shape of the monolithic housing (see below) is universal for SHG to the UV.

\subsection{Length stabilization and its limit}
\label{subsec:locklimit}
During operation, the length of the cavity is stabilized to the PL wavelength via a voltage applied to the piezoelectric actuator using a PI controller. The required error signal is generated using the H{\"a}nsch-Coulliaud (HC) locking scheme\cite{hansch_laser_1980} that generates a dispersion-shaped error signal from the phase shift upon reflection of light from the cavity. P-polarized light is coupled into the cavity and experiences a varying phase shift near the cavity resonance, while s-polarized light does not couple into the cavity and thus obtains a constant $\pi$ phase shift upon reflection. Compared to the Pound-Drever-Hall \cite{drever_laser_1983} locking scheme, no sidebands are imprinted on the PL and thus are also absent in the SHG, which is an advantage for certain applications.

The locking bandwidth is fundamentally limited by the mechanical resonance frequency $f^*$ of the piezoelectric actuator and the attached mirror. It is given as\cite{thorfrequ}
\begin{align}
\label{eq:mres}
f^*=f_0\sqrt{\frac{\frac{m}{3}}{\frac{m}{3}+M}}
\end{align}
where $f_0$ is the mechanical resonance frequency of the piezoelectric actuator, $m$ its mass and $M$ the mass of the attached mirror.

For the monolithic cavity design presented here, we expect only relatively small changes of the round-trip length. Therefore a short piezoelectric actuator with low weight and high resonance frequency seems to be sufficient to keep the cavity length stabilized under moderate environmental perturbations.

\section{Cavity design}
\label{sec:cavdesign}

\begin{figure}
\includegraphics[width=8.0cm]{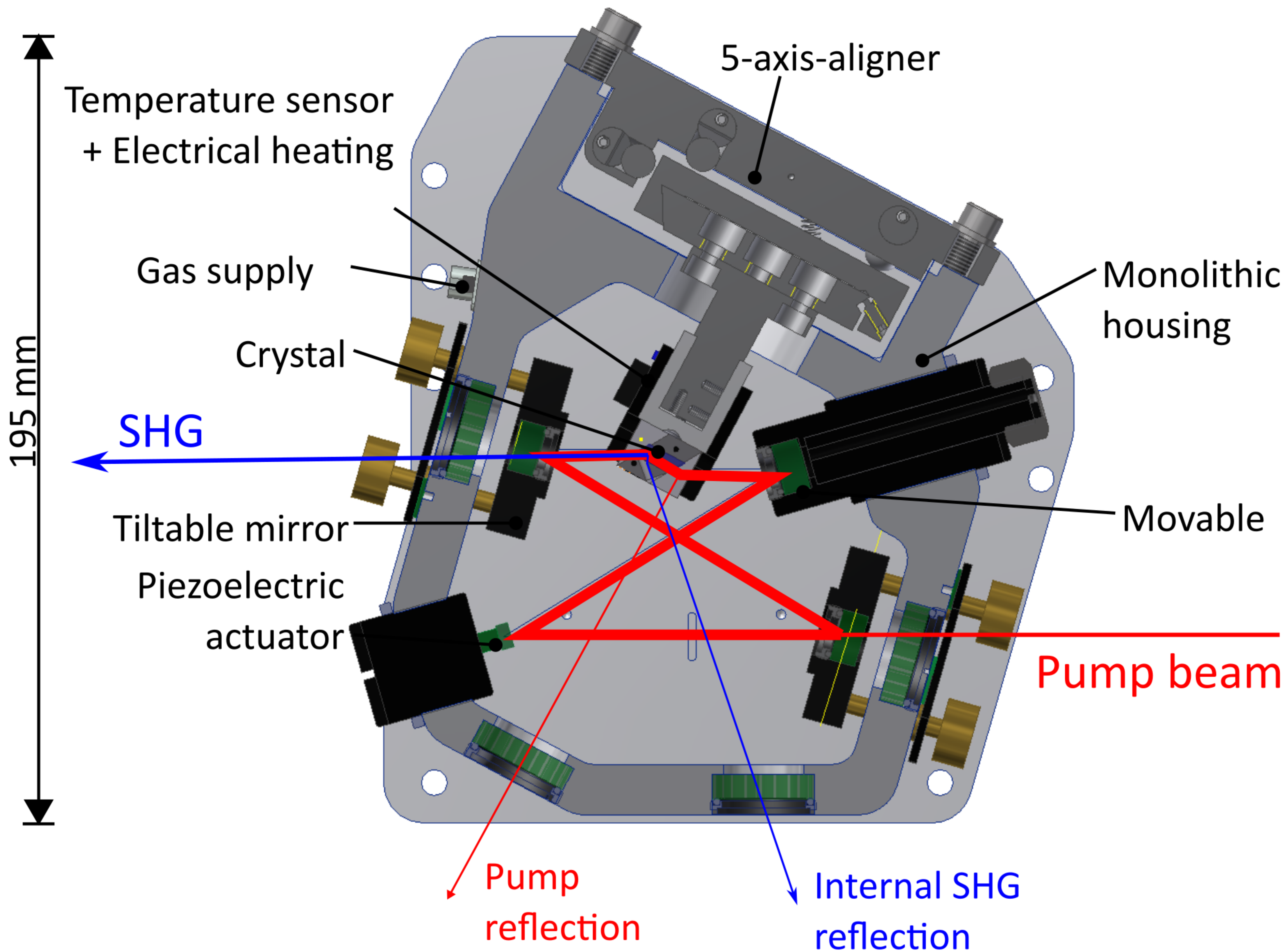}
\caption{Schematic cross-section of the SHG cavity. All mechanical components are mounted directly to the monolithic housing. The incoupling and outcoupling mirror are tiltable using micrometer screws while the other mirrors can be translated only. The temperature-stabilized BBO crystal is mounted on a 5-axis aligner. Ar-coated windows allow for optical access to the sealed cavity.}
\label{fig:shgset}
\end{figure}

The main design goals of the cavity are "turn-key" operation after being exposed to mechanical environmental conditions typical for long distance truck and plane transportation. Furthermore, robust locking and operation in a non-lab environment is important to be a reliable part of a transportable quantum optics experiment.

Therefore, the main body is milled from a single Al block shown in Fig.~\ref{fig:shgset}. Two mirror holder front plates mounted via three micrometer screws ($150\,\mu\mathrm{m}$ displacement per turn) and two springs each, all mounted directly on the main body, feature the four degrees of freedom (DOF) required to close the beampath. In combination with the two other mirrors that are mounted on fine-threaded aluminum cylinders, the cavity round trip length can be adjusted on the few mm-scale while keeping the angles between the beams and therefore the astigmatism compensation constant. The mirrors are mounted using stress-free retaining rings to avoid birefringence caused by mechanical stress. This does not apply for the mirror behind the incoupler which is glued (Thorlabs 353NDPK Epoxy) on a piezoelectric actuator (Thorlabs AE0505D08F) to enable cavity length stabilization.
In crystals used for SHG into the UV, crystal inhomogeneities and degradation effects have been observed. Compensation of these effects requires two translational DOF parallel to the front surface of the crystal. With a third DOF the waist position in the crystal along the direction of the PL propagation is adjusted. Two additional rotational DOF allow the fine adjustment of the phasematching and Brewster's angle. All these DOF are provided by a commercially available 5-axis-aligner (NEWPORT 9081-M) that is mounted directly in the main body. The crystal is mounted in two shells that are mounted to a lever connected to the aligner. These shells and the lever are equipped with channels that lead oxygen from a supply connected to the main body directly to the two facets of the crystal to prevent light-induced damage \cite{JCS1993}. Additionally, the tip of the lever can be activly temperature stabilized using a thick-film resistor (Vishay Sfernice RTO 20) and a sensor (TDK B57861S) connected via a standard D-sub feedthrough to a suitable PI controller to prevent condensate formation on the crystal facets that would likely cause degradation of hygroscopic crystals like BBO. Finally, the cavity is equipped with FKM sealings to prevent unwanted substances from entering the enhancement cavity and cause degradation of the crystal and/or mirrors. Therefore, a gas outlet in the main body ensures a moderate gas exchange rate and overpressure inside the housing. The oxygen concentration can be measured by a sensor (Greisinger GOX 100) located in the lid of the main body.

\section{Setup}
\label{sec:setup}

\begin{figure}[htbp!]
\includegraphics[width=8.0cm]{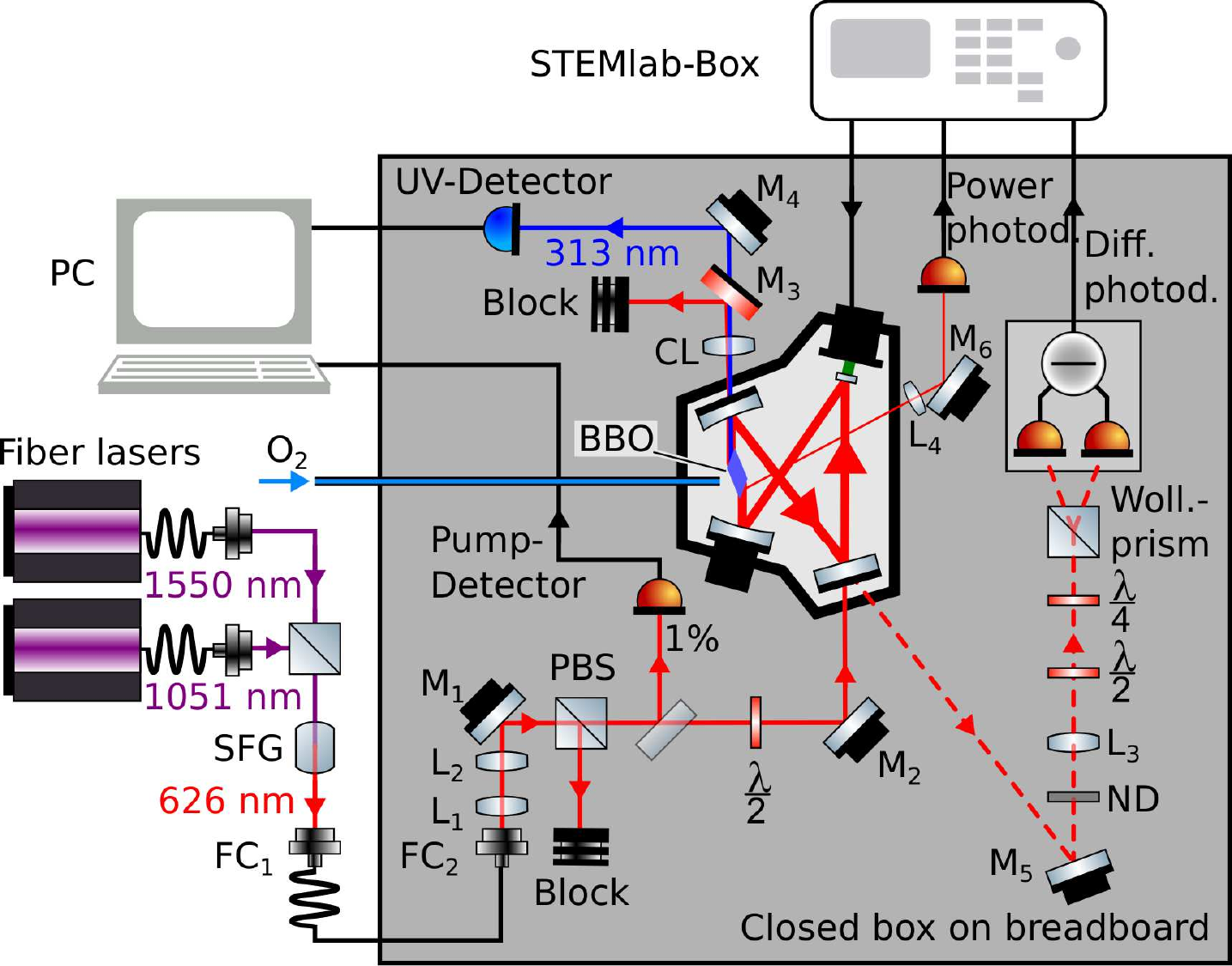}
\caption{Schematic overview of the complete setup for 313 nm generation. $\mathrm{FC}_i:$ fiber coupler, $\mathrm{L}_i:$ lens, $\mathrm{M}i:$ mirror, $\mathrm{PBS}:$ polarizing beam splitter cube, $\frac{\lambda}{2}$/$\frac{\lambda}{4}$ half/quarter wave plate, ND: neutral density filter. 
Left: pump light generation at 626~nm by SFG of two IR fiber lasers. Center: 313~nm generation in the resonant enhancement bow-tie cavity locked by the H{\"a}nsch-Coulliaud locking scheme (dashed beam path and digital multi purpose PI controller STEMlab-Box). For output power stability monitoring, a fraction of the pump and SHG light is picked off and measured via photo diodes.}
\label{fig:comeset}
\end{figure}

Fig.~\ref{fig:comeset} shows a schematic overview of the complete setup for SHG at $313\,\mathrm{nm}$. It consists of a sum frequency generation (SFG) setup for the $626\,\mathrm{nm}$ pumplight, the SHG cavity setup, the locking electronics, and a computer for power recording. The general approach is similar to Wilson et al.\cite{wilson_750-mw_2011}.

\subsection{Pumplight setup}
\label{subsec:pump}
Up to $1.1\,\mathrm{W}$ pump light (PL) at 626 nm are generated by sum frequency generation (SFG) of two fiber lasers at $1550\,\mathrm{nm}$ and $1051\,\mathrm{nm}$ in a PPLN crystal described in\cite{AI2016} and coupled in a polarization maintaining fiber at FC1. This fiber leads to a $12.7\,\mathrm{mm}$ thick $(450\,\mathrm{mm})^2$ aluminum breadboard with all the optical components for the $313\,\mathrm{nm}$ SHG generation. This setup as well as the aforementioned SFG are covered by two boxes made of aluminum. Both are mounted on a standard air damped optical table in a laboratory temperature stabilized at the $1\,\mathrm{K}$ level. The breadboard holding the cavity is placed on $5\,\mathrm{mm}$ thick viscoelastic damping rubber (Sorbothane\textsuperscript{\textregistered}).

\subsection{SHG setup}
\label{subsec:shgsetup}

Behind the PL fiber coupler FC2 a telescope consisting of two lenses L1 and L2 with focal lengths $f=+50\,\mathrm{mm}$ and $f=+30\,\mathrm{mm}$, respectively, mounted on single axis translation stages is used for the modematching with the PL cavity mode. A $1\%$ pickup beam is used to monitor the PL power. The two highly stable mirrors M1 and M2 provide the four DOF required for the coupling the PL into the cavity. Both mirrors are mounted using retaining rings to avoid birefringence due to mechanical stress. A polarizing beamsplitter cube (PBS) cleans the polarization of the PL and a subsequent $\lambda/2$ waveplate rotates it into the plane required for the SHG.
Behind the outcoupling mirror of the SHG cavity a cylindrical lens CL with focal length $f=+100\,\mathrm{mm}$ is employed for correcting the astigmatism arising from the walk-off of the SHG light generated inside the BBO crystal. A second mirror (M3) of the same type as the outcoupler is used to filter the transmitted PL from SHG beam. For the test presented here, a thermal powermeter measurement head is used to collect the SHG light.
The PL reflected by the cavity is attenuated by a 3.0 absorptive neutral density filter (ND), focussed by lens L3 with focal length $f=+250\,\mathrm{mm}$ and send through a polarization analyzer consisting of a half-wave plate, a quarter-wave plate, and a Wollaston prism to generate two orthorgonally polarized beams of similar intensity carrying the information about the phase relation of the PL entering the cavity \cite{hansch_laser_1980}. These two beams hit a differential photodiode. The resulting signal is the error signal for the digital PI controller. The small fraction of PL reflected on the front surface of the crystal due to imperfect polarization and scatter is focused on a photodiode to generate a signal as a measure for the circulating power.

\subsection{Digital PI controller design}
\label{subsec:rp-section}
The error signal generated by the differential photodiode is fed into a digital PI controller based on a modified STEMlab 125-14 (formerly: Red Pitaya). A real-time digital PID controller algorithm is implemented in the FPGA as part of the hardware conﬁguration. It operates independently from the embedded software on the STEMlab. The parameters of the digital controller are set either through a remote network connection, or using a standalone user interface including a touchscreen and rotary encoder inputs. For the purpose of cavity locking, the D-part of the controller is set to zero. Besides the PID controller, a function generator was implemented. It can output sine, square or triangular waveforms within frequency range of 0 - $50\,\mathrm{MHz}$.

\subsubsection{Hardware}
\begin{figure}[htbp!]
\includegraphics[width=8.0cm]{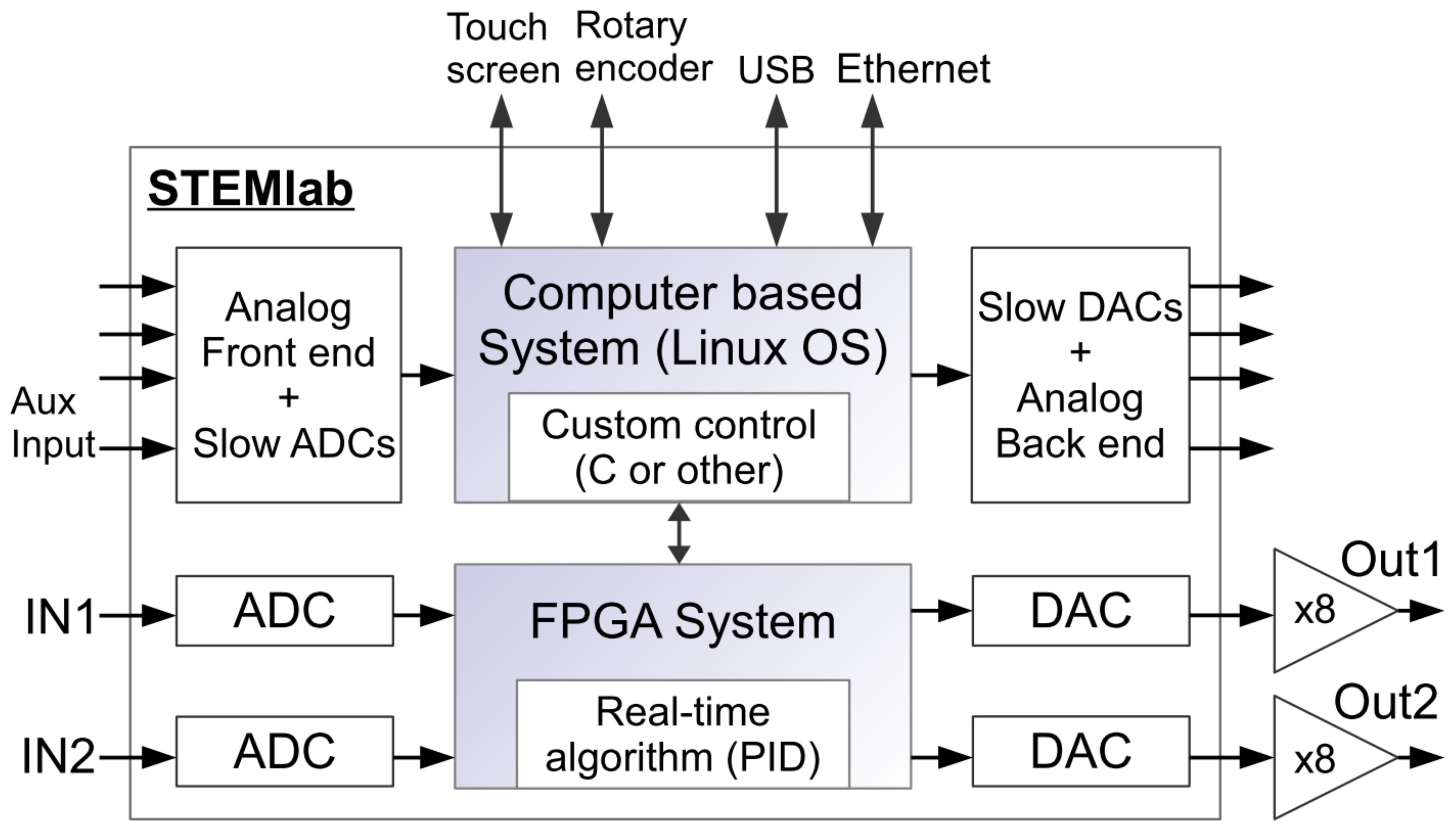}
\caption{Component overview of the STEMlab box cf. \cite{kick_starter}.}
\label{fig:comp_overview}
\end{figure}

STEMlab \cite{rp} hardware is not open-source, but its processing core and peripherals give many options for low-level customization. The hardware platform is built around the Xilinx Zynq7010 All Programmable System-on-Chip (SoC) \cite{Zynq-data,Zynq}, which combines a dual-core ARM Cortex A9 processor, and a field programmable   gate   array   (FPGA)   in   one   chip.   Its   peripherals   include   16   general-purpose input/output (GPIO) lines, various digital interfaces such as I2C, SPI and UART, and four 12-bit ADCs and DACs operating at 100 kS/s. All of these signal lines are accessible on the extension connectors of the board. In addition, there are two high-speed communication interfaces based on Ethernet and USB. The STEMlab provides two fast analog input and output channels, operating at sampling rate of 125 MS/s and having 14-bit nominal resolution. They are implemented using fast ADCs (LTC2145) \cite{ADC} and DACs (AD9767) \cite{DAC}, plus various analog signal conditioning circuits. Fig.~\ref{fig:comp_overview} shows an overview of the components in the STEMlab system that are most relevant for the implemented digital controller. The blue boxes indicate the two principal domains of the Zynq, where custom hardware and software is implemented.

The main characteristics of the fast analog frontend channels are given in Tab. \ref{Tab_technical_data}. They are the most  critical interfaces, because their conversion errors determine the achievable performance of the digital controller. Apart from the effects of quantization that would be present even in ideal converters, various other error sources in the ADCs and DACs have to be considered. Among them are static gain and offset errors, as well as temperature-dependent drifts.

To obtain a stand-alone controller unit based on the STEMlab, we added a user interface consisting of an LCD display with touchscreen and a rotary encoder. The voltage range of the DAC outputs was increased using non-inverting amplifiers with optional external low-pass filters for control applications with reduced bandwidth. Using ampliﬁers with Gain G=8, the output range of $2\,\mathrm{V}_\mathrm{pp}$  was scaled to a $16\,\mathrm{V}_\mathrm{pp}$, sufficient to cover two free spectral ranges of the monolithic doubling cavity, when applied to the cavity piezo-electric actuator. To reduce the output noise of the STEMLab, we implemented the modification described in \cite{LNEUHAUS}, which eliminates the excess noise coupling from the digital supply rail to the analog output. This improvement led to a noise reduction from 3-4 LSB peak-to-peak to 1-2 LSB peak-to-peak. A certain variation of excess noise was observed between different units. In addition to the described circuits, the standalone box contained separate power supplies for the STEMlab board, its cooling fan, and the output amplifiers.  The piezoelectric actuator inside the cavity was connected with an series resistor of $10\,\Omega$, which, together with the capacitance of the actuator, forms a low-pass filter with a cut-off frequency of 21.2~kHz, preventing noise-induced oscillations at high frequency.

\subsubsection{Software and FPGA configuration}
The STEMlab embedded system uses the Linux operating system (presently the Ubuntu 16.04 distribution). The OS image is loaded from an SD card, which also contains the customized software and the FPGA configuration.

The open-source web applications provided with the STEMlab, as well as the PID controller application, can be accessed from any web browser via a network connection. The digitized input signals can be monitored in real time. All parameters of the digital PID controller can be set remotely through the web application, via serial connection, or using the local user interface when operating in standalone mode.

To achieve low-latency digital control, the calculations for the PID algorithm are implemented in the FPGA hardware. Hard real-time processing is possible due to the deterministic timing behavior of the FPGA. The original STEMlab PID application contains a “Multiple Input Multiple Output (MIMO) PID controller which consists of four standard PID controllers with P, I and D parameter settings and integrator reset control” \cite{rpwiki_pid}. The PID parameters can be set in the user interface, and are then sent to the FPGA registers described in the register map \cite{rp_mmap}. The implementation of the four discrete PID controllers can be approximately described by the equations for a continuous-time control system:
\begin{align}
m(t)\,&=\,K_{\mathrm{P}}\cdot e(t)\,+\,K_{\mathrm{I}}\,\int{e(t)\,\mathrm{d}t}\, + \,K_{\mathrm{D}}\cdot \frac{\mathrm{d}e(t)}{\mathrm{d}t} \label{eq:depid}\\
G(s)\,&=\,K_{\mathrm{P}}\,+\,\frac{K_{\mathrm{I}}}{s}\, + \,K_{\mathrm{D}}s, \label{eq:depidtf}
\end{align}
where $m(t)$ is the actuating variable, $e(t)$ the error signal, and $G(s)$ the transfer function of the controller. $K_\mathrm{P}$, $K_\mathrm{I}$, and $K_\mathrm{D}$ are the scaling factors for the  P, I, and D part of the controller. All three summands are calculated in parallel.

\begin{figure}[htbp!]
\includegraphics[width=8.0cm]{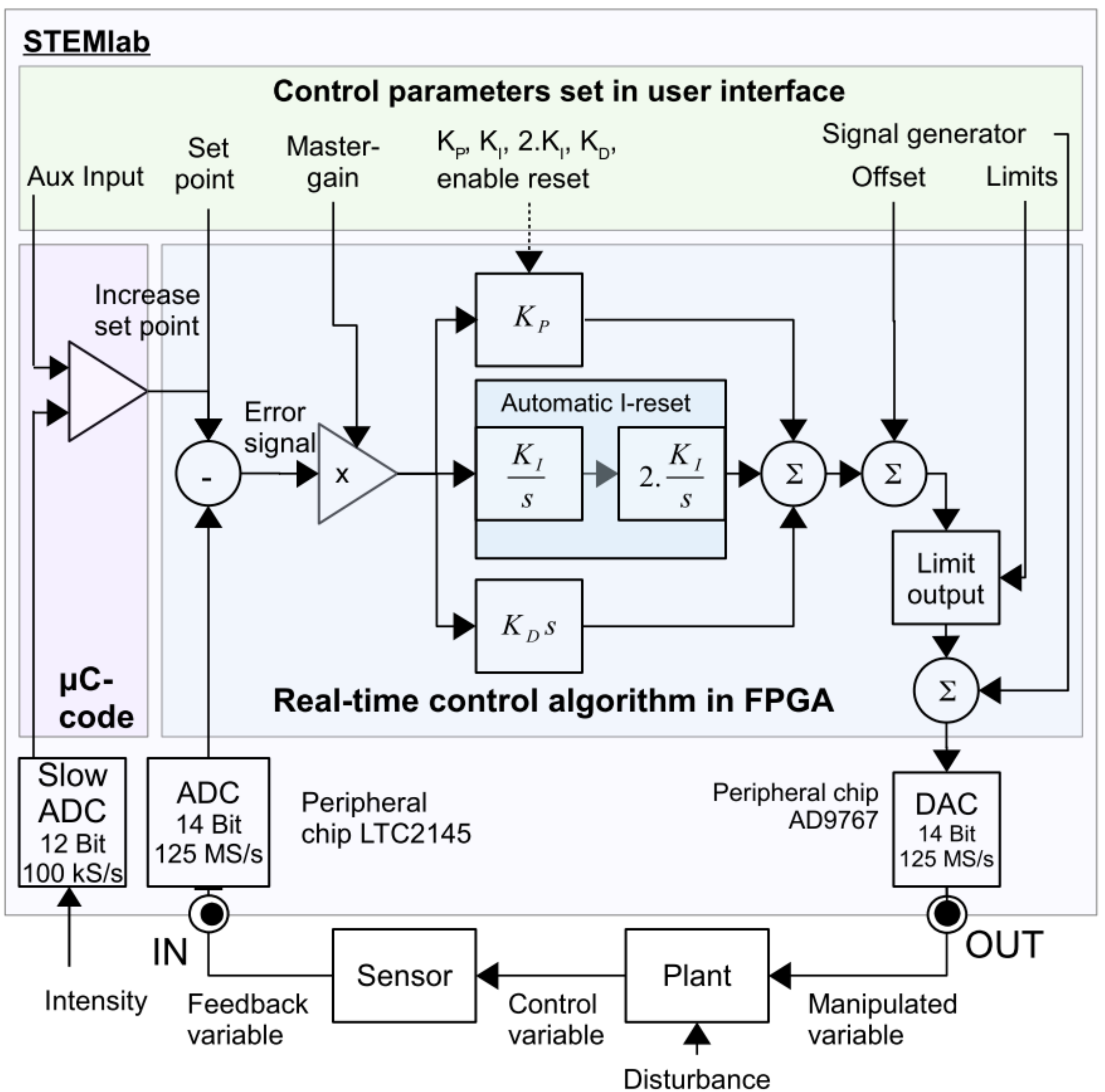}
\caption{Block diagram of the modified STEMlab - PID in a feedback control loop}
\label{fig:RP_scheme}
\end{figure}

To match the specific requirements for quantum optics experiments, the standard PID controller was modified. The source code of the new implementation is free and can be downloaded from \cite{github}. Its  structure is shown in Fig.~\ref{fig:RP_scheme}. A master gain term, a second integral part, and an offset setting were added to the basic PID. The ranges of the control parameters were extended, and the original STEMlab signal generator application was integrated in the controller. 

A number of additional functions were implemented: the settings can be saved, the output voltage can be limited digitally, and an auxiliary input analog signal can be used to enable the controller. The latter function was implemented to allow automatic re-locking of the controller. For this the auxiliary “intensity” signal from the photodetector "Power diode" in Fig.~\ref{fig:comeset} is used to discriminate between actual resonances of the PL TEM$_{00}$-mode and parasitic signals.
Moreover, a automatic integrator reset and a sample-and-hold mode were implemented. All additional functions are optional and run in parallel with the PID algorithm, without affecting its real-time performance.

To allow finer adjustment of the integrator settings, an optional decimation scheme was implemented, in which only every  $N^\mathrm{th}$ sample ($N = 2$ to 1000) is summed.

\subsection{Characteristics of the digital controller}

\subsubsection{Amplitude frequency response}
\begin{figure}[htbp!]
\includegraphics[width=8.0cm]{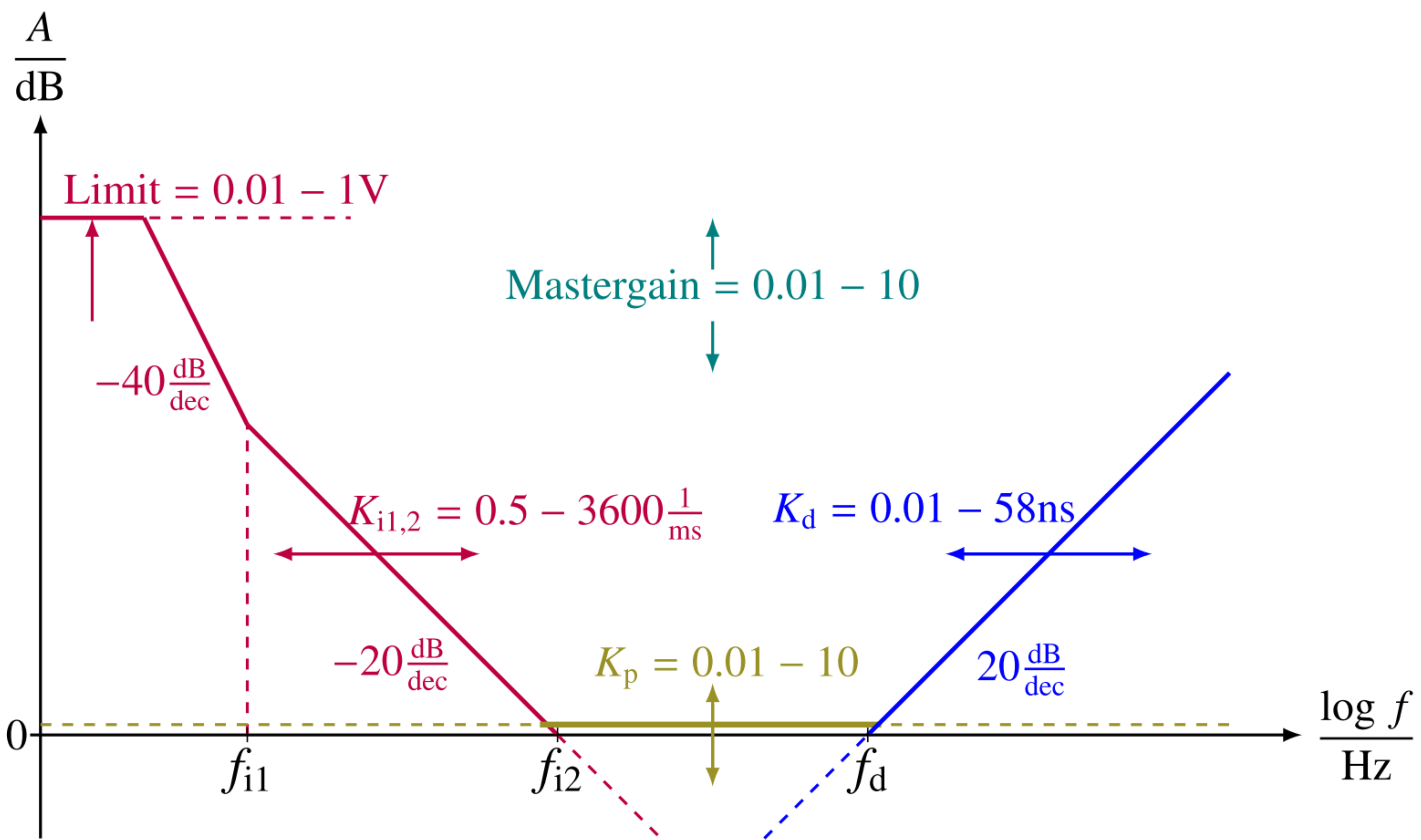}
\caption{Amplitude response of the PID versus frequency. The ranges for the different control parameters are shown together with their effect on the response curve.}
\label{fig:amp_pid_pdh}
\end{figure}

Fig.~\ref{fig:amp_pid_pdh} shows the amplitude response with all ranges and step sizes of the control parameters, as well
as their characteristic frequencies and slopes. The individually adjustable control parameters are the 
master gain G, the P-, I-, $2^{\mathrm{nd}}$  I- and D- contributions of the control loop. 

The physical meaning of the control parameters for the I- and D-parts as $0\,\mathrm{dB}$ gain crossover frequencies 
becomes evident when we convert from angular to linear frequencies (Eq. \eqref{eq:gcfi} and \eqref{eq:gcfd}):
\begin{align}
f_{\mathrm{i}} = \frac{K_{\mathrm{i}}}{2\pi} \label{eq:gcfi} \\
f_{\mathrm{d}} = \frac{1}{2\pi\cdot K_{\mathrm{d}}} \label{eq:gcfd}
\end{align}

\subsubsection{Limits}
\begin{figure}[htb!]
\includegraphics[width=8.0cm]{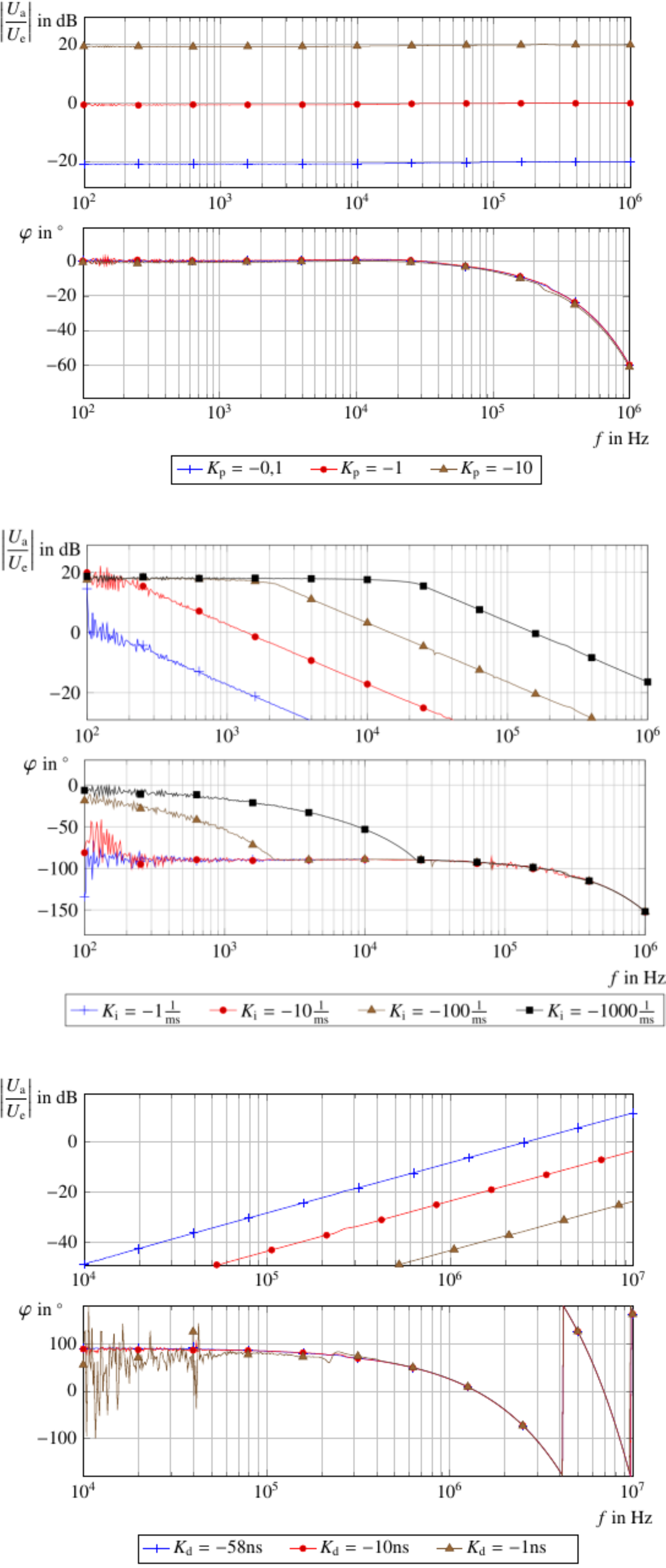}
\caption{Bode diagramm open loop. From top to bottom: proportional/integral/differential part.}
\label{fig:Bode_diagramm}
\end{figure}

The closed-loop bandwidth of the digital controller is fundamentally limited by its group delay $\tau = 155\,\mathrm{ns}$. 
The phase shift from the delay increases linearly with frequency, and cannot be compensated in any real-time causal system. Its effect is visible in the Bode plots of Fig.~\ref{fig:Bode_diagramm}. The additional phase shift  is approximately  $−55.4^\circ$ at $1\,\mathrm{MHz}$, measured in-loop. This suggests a maximum closed-loop control bandwidth of a few MHz, determined by the total phase shift in the loop.

Another limitation exists due to the dynamic range bottlenecks at the ADC and the DAC. The effective 
resolution of the converters is lower than their nominal resolution of 14 bits. This was partially explained 
earlier for the DAC, where an excess digital noise contribution was observed. Fig.~\ref{fig:RP_noise} in the Appendix shows the 
voltage noise spectral density measured at the output of two STEMlab boards, and two standalone boxes 
with and without an output amplifier. The integrated noise voltage within $1\,\mathrm{Hz}$ to $1\,\mathrm{MHz}$ is measured to be $76.89 \, \mathrm{\mu V}$ for the unmodified STEMlab board without amplifier stage and deactivated PID controller (A, red curve). With the offset resistors removed, $32.44 \, \mathrm{\mu V}$ is measured (B, black curve). For the modified STEMlab installed inside the standalone box with activated PID controller, $120.12 \, \mathrm{\mu V}$ is obtained (C, green curve). Finally, an integrated noise voltage of $392.66 \, \mathrm{\mu V}$ is measured for the STEMlab board with offset resistors removed inside the standalone box and with amplifier stage added.

\section{Experiment}
\label{sec:experiment}
During all experiments the cavity was flushed with $99.95\%$ pure oxygen at a rate of $0.022\,\mathrm{l/min}$.
A finesse $F_\mathrm{meas}=317$ was measured with the Brewster-cut BBO crystal in place and mirrors with reflectivities of $R_1=98.50\%$ and $R_2=R_3=R_4=99.95\%$.

\subsection{Long term stability}
\label{subsec:ltlock}
\begin{figure*}[htb!]
\includegraphics[width=16.0cm]{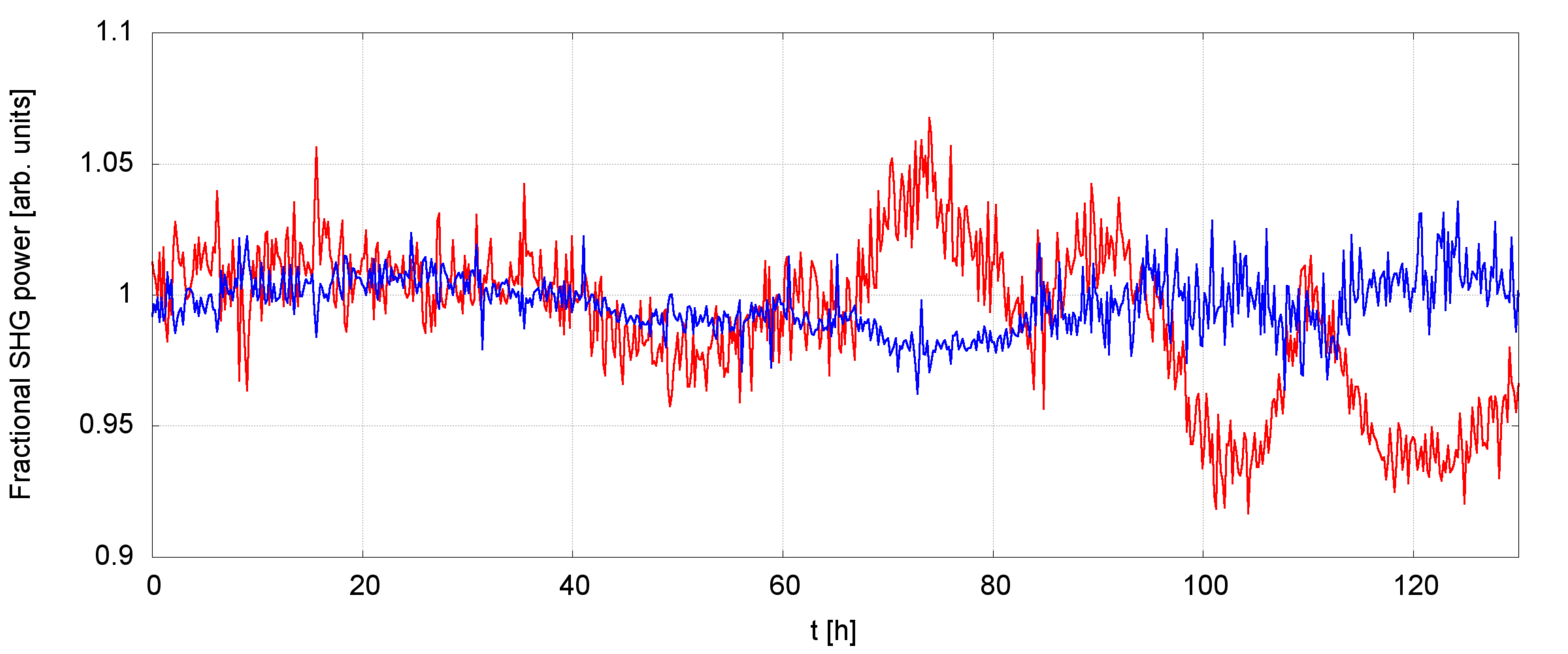}
\caption{Long term SHG output power measurement. Red: measured SHG power, a power of "1" equals $186\,\mathrm{mW}$. Blue: SHG normalized to the PL power squared as a measure of the conversion efficiency.}
\label{fig:ltlock}
\end{figure*}

For many applications in quantum optics a powerstable UV source is required that reliably stays in lock over many hours. In order to demonstrate that our setup fulfills those requirements, it has been locked over $130\,\mathrm{h}$ during which the PL power and SHG Power have been recorded every $1\,\mathrm{s}$ using two power meters (Thorlabs PM100D with S130VC Sensor and Ophir Vega with PD300-UV Sensor, cf. Fig.~\ref{fig:comeset}). Fig.~ \ref{fig:ltlock} shows the measured SHG output power scaled to the average SHG power (red) and to the instantaneous PL power (blue). The SHG power stayed on a level of about $186\,\mathrm{mW}$, while the cavity was pumped with about $530\,\mathrm{mW}$ PL. The SHG power fluctuates by about $\pm7\%$ peak-to-peak ($3\%$ rms) around its mean value on a few hour timescale, while the SHG power rescaled to the PL varies only by $\pm3\%$ peak-to-peak ($1\%$ rms). This suggests that the SHG power fluctuations were mainly caused by pump power fluctuations, indicating that the alignment of the cavity did not change on relevant length or angular scales over the duration of the measurement.

To our knowledge, this is the longest published in-lock operation of a resonant SHG ring cavity. The output power of the SHG cavity fully agrees with the corresponding measurement and calculation presented in \cite{wilson_750-mw_2011}.

\subsection{Lock performance}
\label{subsec:pidnoi}
\begin{figure*}[htbp!]
\includegraphics[width=16.0cm]{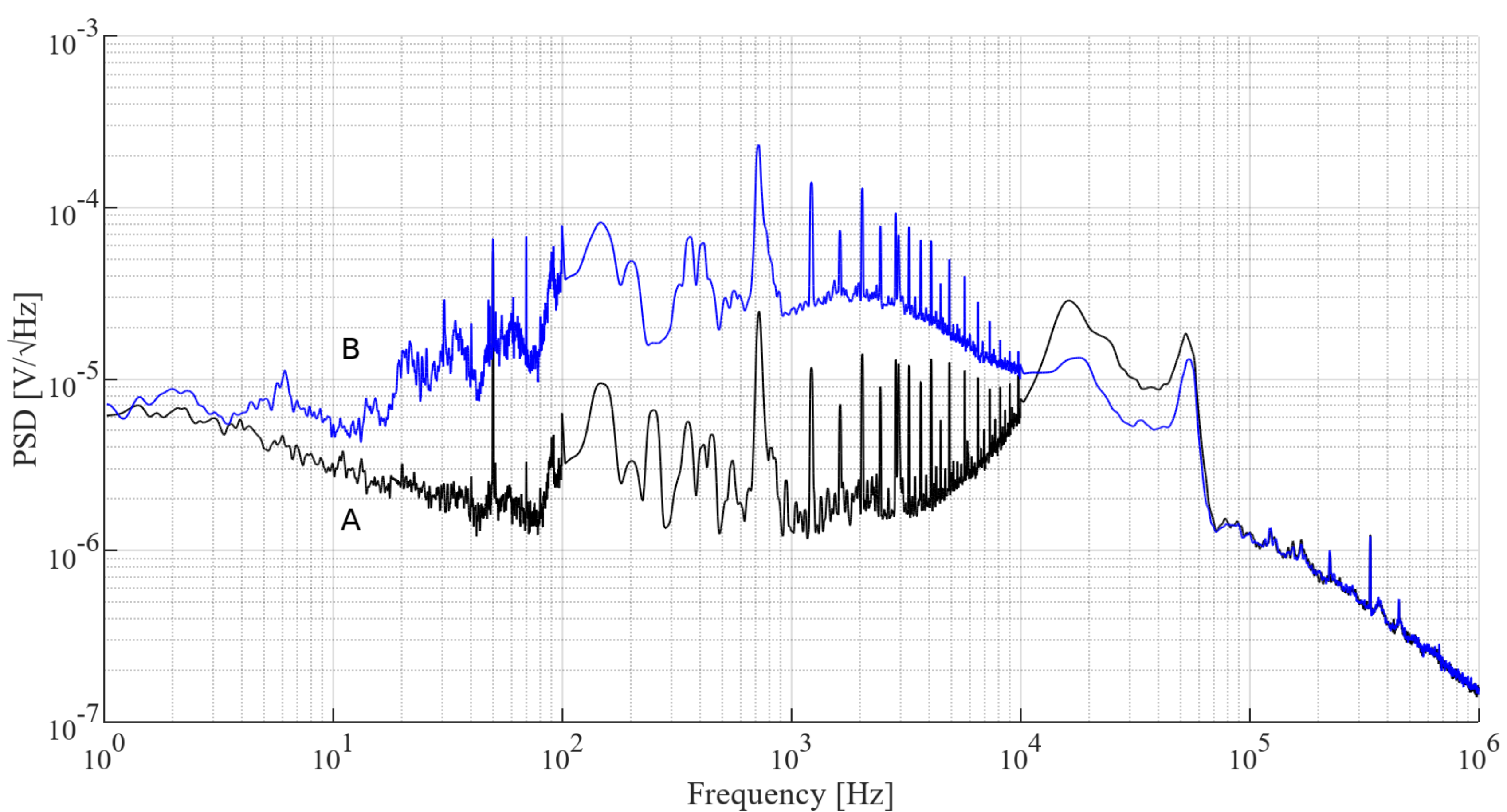}
\caption{Error signal (=STEMlab input) power spectral density for two different PI settings. A (black): lock with optimized gain settings, B (blue): low-gain lock.}
\label{fig:RP_noise}
\end{figure*}

The performance of the cavity lock has been investigated by recording the power spectral density of the error signal generated by the differential photo diode (s. Fig.~\ref{fig:comeset}) using a vector signal analyzer (Agilent 89441A). Fig.~\ref{fig:RP_noise} shows the error signal for a loose lock achieved by choosing gain setting just sufficient enough to lock (blue) and for optimized settings (black). With the optimized lock a suppression by up to almost one order of magnitude for acoustic frequencies is achieved. A broad resonance at about $17\,\mathrm{kHz}$ sets the upper limit for the locking bandwidth in this setup. This frequency is likely to be the first mechanical resonance of the small mirror, the piezoelectric actuator, and its holder. In a future setup, it could possibly be improved by applying the mount design presented in \cite{Jitschin_fast_1979,briles_simple_2010,goldovsky_simple_2016}, since Equ.~\ref{eq:mres} results in an upper limit of $f^*\approx 120 \,\mathrm{kHz}$ for the piezo electric actuator and mirror used here.

\subsection{Acceleration sensitivity tests}
\label{subsec:vstest}
\begin{figure}[htbp!]
\includegraphics[width=8.0cm]{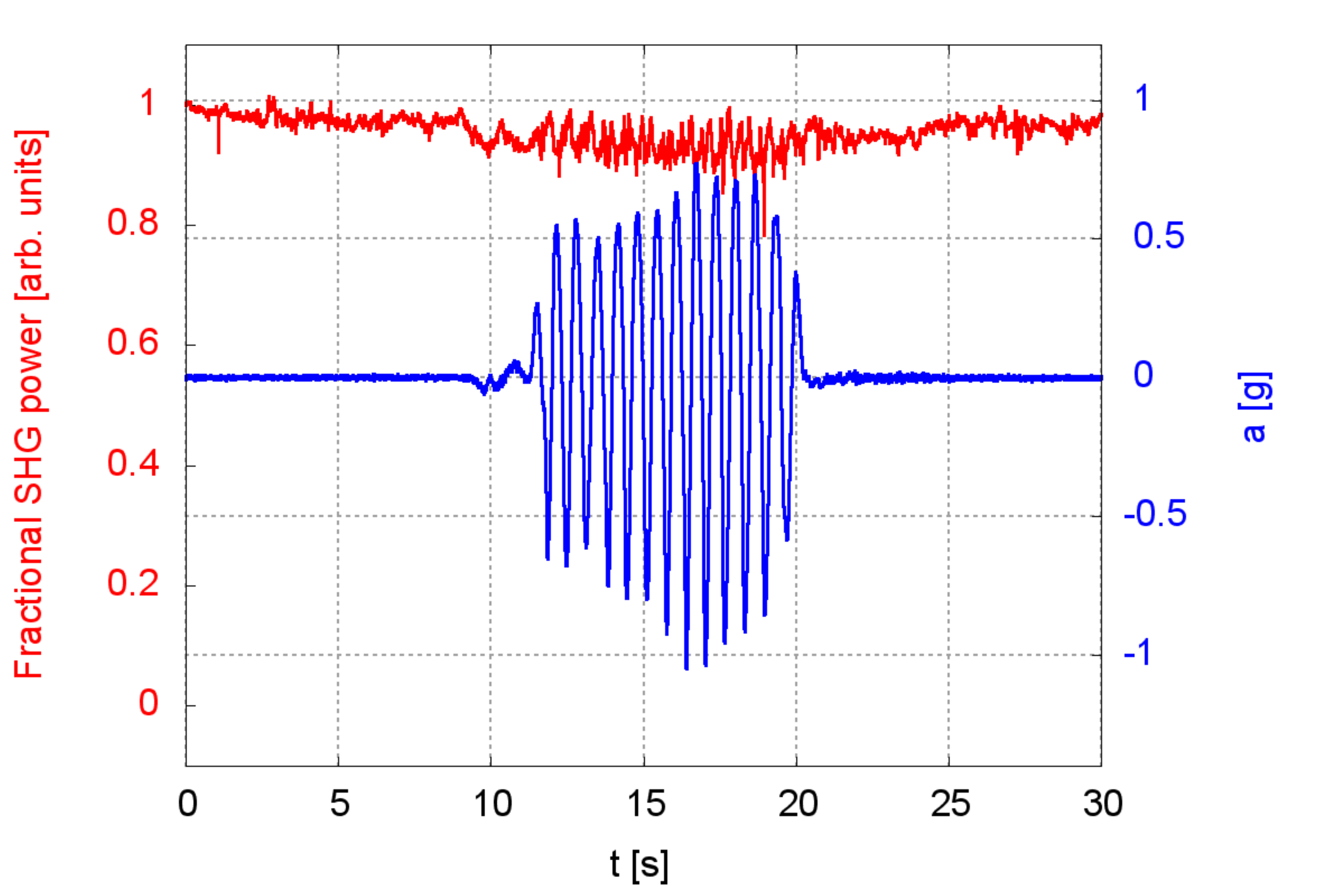}
\caption{Cavity SHG output power and acceleration versus time during acceleration excitation in vertical direction. Red: SHG power, blue: vertical acceleration. For accelerations up to 1 g (gravity substracted) the SHG output power fluctuates on the $10\%$ level while the cavity stays in lock.}
\label{fig:man_shake}
\end{figure}

The monolithic cavity described here is a prototype to be used for generating the clock transition probe light of a transportable $\mathrm{Al}^+$ clock to be used in relativistic geodesy campaigns. In order to prove the suitability of the doubling cavity for such an application, the entire breadboard was accelerated in vertical direction while the cavity was locked. The acceleration was measured using an analog MEMS sensor (ADXL345) glued on the top lid of the cavity. Fig.~\ref{fig:man_shake} shows the output power and vertical acceleration versus time. The measurement shows temporary SHG power drops of about $10\%$ synchronous the acceleration acting on the system. While an absolute acceleration of up to $1\,\mathrm{g}$ was obtained, the cavity stayed in lock and fully recovered the initial SHG power after the acceleration stopped.

\begin{figure}[htbp!]
\includegraphics[width=8.0cm]{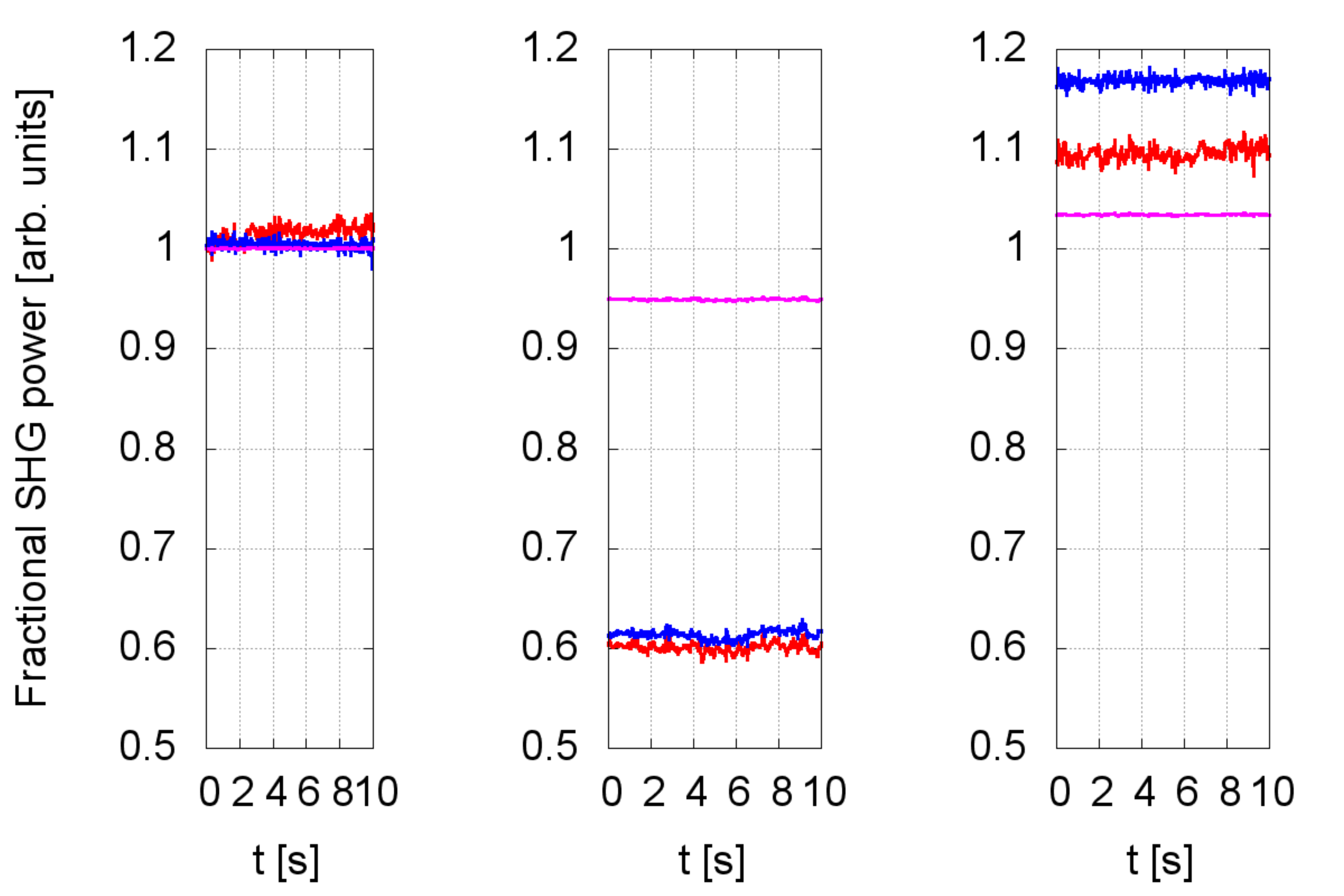}
\caption{SHG power before (left) and right after (middle) being exposed to a 30 min ISO 13355:2016 shaker test. Right: following optimization of the incoupling beam after the shakertest. Red: measured SHG power, blue: SHG normalized to the PL power squared, pink: SHG normalized to the circulating PL power as a measure of the conversion efficiency.}
\label{fig:table_shake}
\end{figure}

In order to demonstrate the reliable operation of the system after transport, it was exposed to an acceleration profile typical for on road transportation. Therefore, the system was optimized first and then the SHG breadboard was placed on a multi-component acceleration exciter \cite{MCAC} and exposed to a vibration power spectrum in the vertical direction according to ISO 13355:2016 with a total acceleration of $0.604\,\mathrm{g}_\mathrm{rms}$ for 30 minutes. Afterwards, the SHG breadboard was reconnected to the rest of the system and locked without further realignment or optimization. The results are shown in Fig.~\ref{fig:table_shake}. The obtained SHG power normalized to the pump power dropped to about $60\%$, while the SHG power normalized to the squared pump power circulating in the cavity remained almost constant. Since this indicates a drop in incoupling efficiency, the alignment of the mirrors M1 and M2 in Fig.~\ref{fig:comp_overview} was optimized, resulting in approximately the same SHG power as before the acceleration exposure test. This strongly indicates that the alignment of the cavity itself was not substantially affected by the shaking test. Therefore the same test was repeated with five times the original acceleration ($3.020\,\mathrm{g}_\mathrm{rms}$), also resulting in the same SHG output power after shaking and alignment correction of the mirrors M1 and M2. From these measurements we conclude that the optical alignment of the cavity itself can withstand typical on road transport situations without any deterioration, while the mechanical stability of the input coupling optics and/or fiber connector for the PL needs to be improved.

\section{Summary and Outlook}
\label{sec:sumout}
A mechanically stable monolithic enhancement cavity for SHG generation in the UV was demonstrated, including in-lock SHG power measurements during acceleration excitation. Less than $10\%$ SHG power reduction during exposure of up to $1\,\mathrm{g}$ were observed with full recovery of the initial SHG power after acceleration stopped. It was shown that the cavity optical alignment can withstand 30 minutes of acceleration excitation with $3.020\,\mathrm{g}_\mathrm{rms}$. This is five times the acceleration amplitude as specified in the corresponding ISO13355:2016 norm, demonstrating the suitability for transportable experiments. 
$130\,\mathrm{h}$ uninterrupted operation without decay in output power at $313\,\mathrm{nm}$ was demonstrated. During this time the SHG power scaled to the pump power fluctuated by $1\%$~rms. The basic design can easily be adapted to other resonator geometries in order to install crystals of different materials or shapes. The locking bandwidth of the setup presented here is $17\,\mathrm{kHz}$. To obtain a higher locking bandwidth in the future, the design of the piezoelectrically actuated mirror mount demonstrated in \cite{Jitschin_fast_1979,briles_simple_2010,goldovsky_simple_2016} could be adopted. To reduce the risk of unwanted substances inside the cavity causing degradation of the crystal and/or mirrors in the next cavity generation, the housing can be left unanodized and metal sealings can be employed. For improved mechanical robustness and leak tightness, the crystal aligner can be made part of the cavity housing.  

\begin{acknowledgments}
We thank L. Klaus for performing the acceleration excitation measurements on the Multi-component acceleration exciter and S. A. King for stimulating discussion.
We acknowledge support from DFG through grants CRC 1128 geo-\textit{Q}, project A03, CRC 1227 DQ-\textit{mat}, projects 
B03 and B06, and Leibniz Gemeinschaft through grant SAW-2013-FBH-3.
\end{acknowledgments}

\newpage 
\section{Appendix}
\label{sec:app1}

\begin{table*} [htbp!]
\caption{STEMlab 125-14 analog front- and backend specifications \cite{rpwiki_h}}
\label{Tab_technical_data}
\begin{small}
\begin{center}
\begin{ruledtabular}
\begin{tabular}{ll|ll} 
\multicolumn{2}{c}{\textbf{Inputs}} & \multicolumn{2}{c}{\textbf{Outputs}}
\tabularnewline
\toprule
Number of channels: & 2 & Quantity: & 2 
\tabularnewline
 Bandwidth: & ${50\,\mathrm{MHz}}$ & Bandwidth: & ${50\,\mathrm{MHz}}$ 
 \tabularnewline
 Sample rate: & ${125\,\mathrm{MS/s}}$ & Sample rate: & ${125\,\mathrm{MS/s}}$ 
 \tabularnewline
 DAC resolution: & ${14\,\mathrm{Bit}}$ & ADC resolution: & ${14\,\mathrm{Bit}}$ 
 \tabularnewline
 \mbox{Full scale voltage:} & selectable: ${\pm\,1\,\mathrm{V}}$, ${\pm\,20\,\mathrm{V}}$ & \mbox{Full scale power:} & ${9\; \mathrm{dBm}}$ 
  \tabularnewline
\mbox{Minimal Voltage Sensitivity:} & ${\pm\,0.122\,\mathrm{mV}}$, ${\pm\,2.44\,\mathrm{mV}}$ & \mbox{1 LSB:} & ${0.122\,\mathrm{mV}}$  
 \tabularnewline
 Input impedance: & ${1\,\mathrm{M\Omega}\; \vert \vert\; 10\,\mathrm{pF}}$ & Load impedance (for full scale power): & ${50\,\mathrm{\Omega}}$ 
 \tabularnewline
 DC offset error: & ${<5\,\mathrm{\%}}$ & DC offset error: & ${<5\,\mathrm{\%}}$ 
 \tabularnewline
  Gain error: & LV: ${<3\,\mathrm{\%}}$, HV: ${<10\,\mathrm{\%}}$ & Gain error: & ${<5\,\mathrm{\%}}$ 
 \tabularnewline
 Input noise level: & ${-119\,\mathrm{dBm/Hz}}$ & Slew rate limit: & ${200\,\frac{\mathrm{V}}{\mu\mathrm{s}}}$ 
 \tabularnewline
\end{tabular}
\end{ruledtabular}
\end{center}
\end{small}
\end{table*}

\begin{table*} [htbp!]
\caption{Limit values of the control parameters}
\label{Tab_params}
\begin{small}
\begin{center}
\begin{ruledtabular}
\begin{tabular}{llllll} 
\textbf{Parameter} &\textbf{Maximum} &\textbf{Step size} &\textbf{Min. frequency} &\textbf{Max. frequency} & \textbf{Inaccuracy}
\tabularnewline
\toprule
 Gain : & ${10}$ & ${0.01}$ & - & - & ${<10\,\%}$
\tabularnewline
 ${K_{\mathrm{p}}}$ : & ${10}$ & ${0.01}$ & - & - & ${<10\,\%}$
 \tabularnewline
 ${K_{\mathrm{i}}}$ : & ${3600\, \frac{1}{\mathrm{ms}}}$ & ${0.5 \,\frac{1}{\mathrm{ms}}}$ & ${80\,\mathrm{Hz}}$ & ${573\,\mathrm{kHz}}$ & ${<17\,\%}$
 \tabularnewline
 ${K_{\mathrm{d}}}$ : & ${58 \,\mathrm{ns}}$ & ${0.01\, \mathrm{ns}}$ & ${2.74\,\mathrm{MHz}}$ & ${16\,\mathrm{GHz}}$ & ${<6\,\%}$
 \tabularnewline
\end{tabular}
\end{ruledtabular}
\end{center}
\end{small}
\end{table*}

\begin{figure*}[htbp!]
\includegraphics[width=17.0cm]{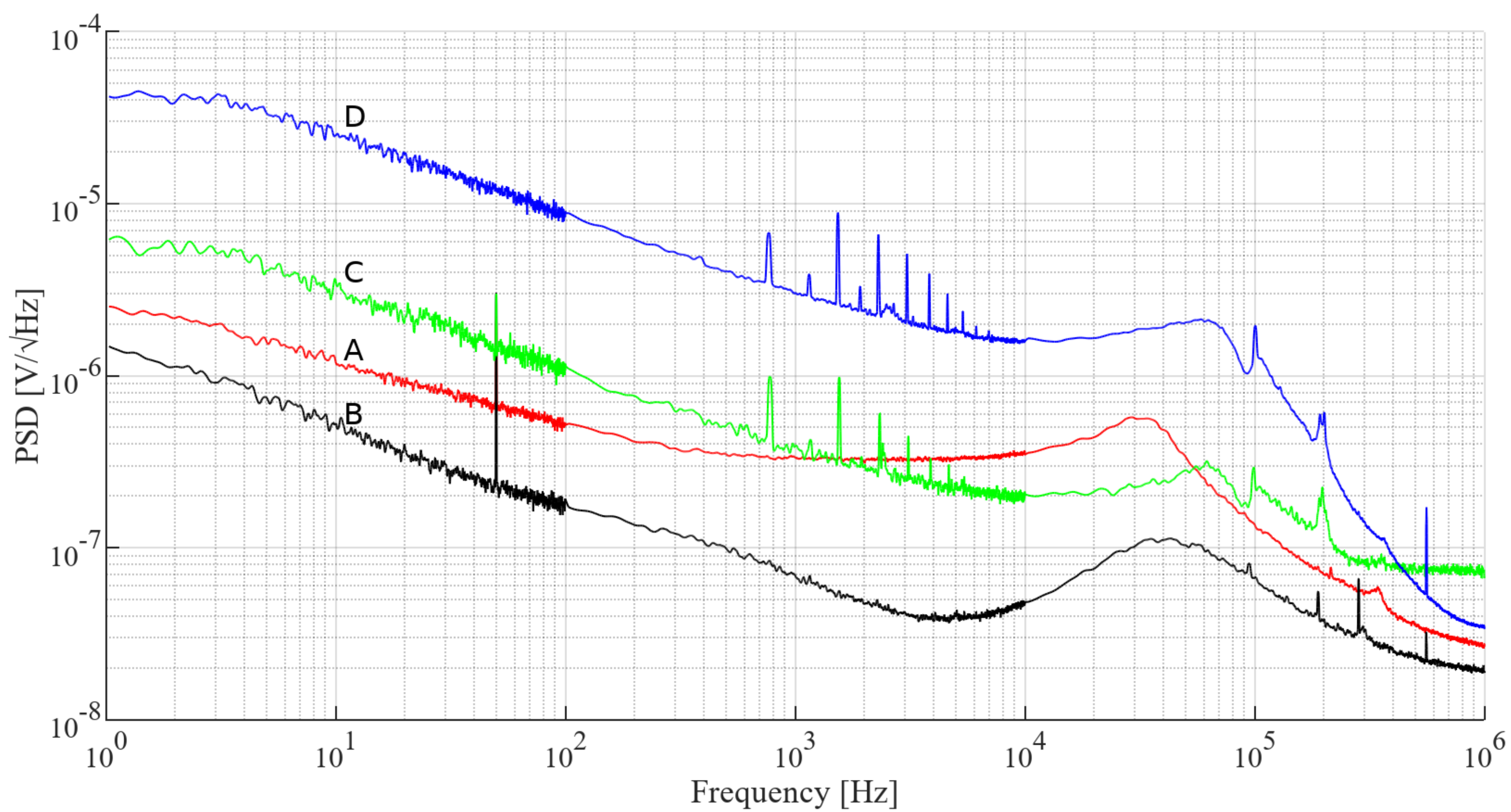}
\caption{Output noise (=STEMlab output) power spectral density with terminated input. A (red): unmodified STEMlab without amplifier stage, PID deactivated. B (black): the same as A, but with offset resistors removed. C (green): STEMlab in the box, PID enabled, without amplifier, $K_{p}=1$. D (blue): the same as C, but with amplifier (8x) and 100~KHz lowpass filter.\\Integrated output voltage noise (bandwidth {$1\,\mathrm{Hz}\,-\,1\,\mathrm{MHz}$}): A = {$76.89 \, \mathrm{\mu V}$}, B = {$32.44 \, \mathrm{\mu V}$}, C = {$120.12 \, \mathrm{\mu V}$}, D = {$392.66 \, \mathrm{\mu V}$}}
\label{fig:RP_out_noise}
\end{figure*}

\pagebreak


\begin{thebibliography}{72}%
\makeatletter
\providecommand \@ifxundefined [1]{%
 \@ifx{#1\undefined}
}%
\providecommand \@ifnum [1]{%
 \ifnum #1\expandafter \@firstoftwo
 \else \expandafter \@secondoftwo
 \fi
}%
\providecommand \@ifx [1]{%
 \ifx #1\expandafter \@firstoftwo
 \else \expandafter \@secondoftwo
 \fi
}%
\providecommand \natexlab [1]{#1}%
\providecommand \enquote  [1]{``#1''}%
\providecommand \bibnamefont  [1]{#1}%
\providecommand \bibfnamefont [1]{#1}%
\providecommand \citenamefont [1]{#1}%
\providecommand \href@noop [0]{\@secondoftwo}%
\providecommand \href [0]{\begingroup \@sanitize@url \@href}%
\providecommand \@href[1]{\@@startlink{#1}\@@href}%
\providecommand \@@href[1]{\endgroup#1\@@endlink}%
\providecommand \@sanitize@url [0]{\catcode `\\12\catcode `\$12\catcode
  `\&12\catcode `\#12\catcode `\^12\catcode `\_12\catcode `\%12\relax}%
\providecommand \@@startlink[1]{}%
\providecommand \@@endlink[0]{}%
\providecommand \url  [0]{\begingroup\@sanitize@url \@url }%
\providecommand \@url [1]{\endgroup\@href {#1}{\urlprefix }}%
\providecommand \urlprefix  [0]{URL }%
\providecommand \Eprint [0]{\href }%
\providecommand \doibase [0]{http://dx.doi.org/}%
\providecommand \selectlanguage [0]{\@gobble}%
\providecommand \bibinfo  [0]{\@secondoftwo}%
\providecommand \bibfield  [0]{\@secondoftwo}%
\providecommand \translation [1]{[#1]}%
\providecommand \BibitemOpen [0]{}%
\providecommand \bibitemStop [0]{}%
\providecommand \bibitemNoStop [0]{.\EOS\space}%
\providecommand \EOS [0]{\spacefactor3000\relax}%
\providecommand \BibitemShut  [1]{\csname bibitem#1\endcsname}%
\let\auto@bib@innerbib\@empty
\bibitem [{\citenamefont {Tan}\ \emph {et~al.}(2015)\citenamefont {Tan},
  \citenamefont {Gaebler}, \citenamefont {Lin}, \citenamefont {Wan},
  \citenamefont {Bowler}, \citenamefont {Leibfried},\ and\ \citenamefont
  {Wineland}}]{tan_multi-element_2015}%
  \BibitemOpen
  \bibfield  {author} {\bibinfo {author} {\bibfnamefont {T.~R.}\ \bibnamefont
  {Tan}}, \bibinfo {author} {\bibfnamefont {J.~P.}\ \bibnamefont {Gaebler}},
  \bibinfo {author} {\bibfnamefont {Y.}~\bibnamefont {Lin}}, \bibinfo {author}
  {\bibfnamefont {Y.}~\bibnamefont {Wan}}, \bibinfo {author} {\bibfnamefont
  {R.}~\bibnamefont {Bowler}}, \bibinfo {author} {\bibfnamefont
  {D.}~\bibnamefont {Leibfried}}, \ and\ \bibinfo {author} {\bibfnamefont
  {D.~J.}\ \bibnamefont {Wineland}},\ }\href {\doibase 10.1038/nature16186}
  {\bibfield  {journal} {\bibinfo  {journal} {Nature}\ }\textbf {\bibinfo
  {volume} {528}},\ \bibinfo {pages} {380} (\bibinfo {year}
  {2015})}\BibitemShut {NoStop}%
\bibitem [{\citenamefont {Chou}\ \emph {et~al.}(2010)\citenamefont {Chou},
  \citenamefont {Hume}, \citenamefont {Koelemeij}, \citenamefont {Wineland},\
  and\ \citenamefont {Rosenband}}]{chou_frequency_2010}%
  \BibitemOpen
  \bibfield  {author} {\bibinfo {author} {\bibfnamefont {C.~W.}\ \bibnamefont
  {Chou}}, \bibinfo {author} {\bibfnamefont {D.~B.}\ \bibnamefont {Hume}},
  \bibinfo {author} {\bibfnamefont {J.~C.~J.}\ \bibnamefont {Koelemeij}},
  \bibinfo {author} {\bibfnamefont {D.~J.}\ \bibnamefont {Wineland}}, \ and\
  \bibinfo {author} {\bibfnamefont {T.}~\bibnamefont {Rosenband}},\ }\href
  {\doibase 10.1103/PhysRevLett.104.070802} {\bibfield  {journal} {\bibinfo
  {journal} {Phys. Rev. Lett.}\ }\textbf {\bibinfo {volume} {104}},\ \bibinfo
  {pages} {070802} (\bibinfo {year} {2010})}\BibitemShut {NoStop}%
\bibitem [{\citenamefont {Yamanaka}\ \emph {et~al.}(2015)\citenamefont
  {Yamanaka}, \citenamefont {Ohmae}, \citenamefont {Ushijima}, \citenamefont
  {Takamoto},\ and\ \citenamefont {Katori}}]{yamanaka_frequency_2015}%
  \BibitemOpen
  \bibfield  {author} {\bibinfo {author} {\bibfnamefont {K.}~\bibnamefont
  {Yamanaka}}, \bibinfo {author} {\bibfnamefont {N.}~\bibnamefont {Ohmae}},
  \bibinfo {author} {\bibfnamefont {I.}~\bibnamefont {Ushijima}}, \bibinfo
  {author} {\bibfnamefont {M.}~\bibnamefont {Takamoto}}, \ and\ \bibinfo
  {author} {\bibfnamefont {H.}~\bibnamefont {Katori}},\ }\href {\doibase
  10.1103/PhysRevLett.114.230801} {\bibfield  {journal} {\bibinfo  {journal}
  {Phys. Rev. Lett.}\ }\textbf {\bibinfo {volume} {114}},\ \bibinfo {pages}
  {230801} (\bibinfo {year} {2015})}\BibitemShut {NoStop}%
\bibitem [{\citenamefont {Hauth}\ \emph {et~al.}(2013)\citenamefont {Hauth},
  \citenamefont {Freier}, \citenamefont {Schkolnik}, \citenamefont {Senger},
  \citenamefont {Schmidt},\ and\ \citenamefont {Peters}}]{hauth_first_2013}%
  \BibitemOpen
  \bibfield  {author} {\bibinfo {author} {\bibfnamefont {M.}~\bibnamefont
  {Hauth}}, \bibinfo {author} {\bibfnamefont {C.}~\bibnamefont {Freier}},
  \bibinfo {author} {\bibfnamefont {V.}~\bibnamefont {Schkolnik}}, \bibinfo
  {author} {\bibfnamefont {A.}~\bibnamefont {Senger}}, \bibinfo {author}
  {\bibfnamefont {M.}~\bibnamefont {Schmidt}}, \ and\ \bibinfo {author}
  {\bibfnamefont {A.}~\bibnamefont {Peters}},\ }\href {\doibase
  10.1007/s00340-013-5413-6} {\bibfield  {journal} {\bibinfo  {journal} {Appl.
  Phys. B.}\ }\textbf {\bibinfo {volume} {113}},\ \bibinfo {pages} {49}
  (\bibinfo {year} {2013})}\BibitemShut {NoStop}%
\bibitem [{\citenamefont {Barrett}\ \emph {et~al.}(2014)\citenamefont
  {Barrett}, \citenamefont {Gominet}, \citenamefont {Cantin}, \citenamefont
  {Antoni-Micollier}, \citenamefont {Bertoldi}, \citenamefont {Battelier},
  \citenamefont {Bouyer}, \citenamefont {Lautier},\ and\ \citenamefont
  {Landragin}}]{barrett_mobile_2013}%
  \BibitemOpen
  \bibfield  {author} {\bibinfo {author} {\bibfnamefont {B.}~\bibnamefont
  {Barrett}}, \bibinfo {author} {\bibfnamefont {P.-A.}\ \bibnamefont
  {Gominet}}, \bibinfo {author} {\bibfnamefont {E.}~\bibnamefont {Cantin}},
  \bibinfo {author} {\bibfnamefont {L.}~\bibnamefont {Antoni-Micollier}},
  \bibinfo {author} {\bibfnamefont {A.}~\bibnamefont {Bertoldi}}, \bibinfo
  {author} {\bibfnamefont {B.}~\bibnamefont {Battelier}}, \bibinfo {author}
  {\bibfnamefont {P.}~\bibnamefont {Bouyer}}, \bibinfo {author} {\bibfnamefont
  {J.}~\bibnamefont {Lautier}}, \ and\ \bibinfo {author} {\bibfnamefont
  {A.}~\bibnamefont {Landragin}},\ }\href {http://arxiv.org/abs/1311.7033}
  {\bibfield  {journal} {\bibinfo  {journal} {Proc. Int'l Sch. Phys. ‘Enrico
  Fermi’}\ ,\ \bibinfo {pages} {493}} (\bibinfo {year} {2014})},\ \bibinfo
  {note} {arXiv: 1311.7033}\BibitemShut {NoStop}%
\bibitem [{\citenamefont {Huntemann}\ \emph {et~al.}(2016)\citenamefont
  {Huntemann}, \citenamefont {Sanner}, \citenamefont {Lipphardt}, \citenamefont
  {Tamm},\ and\ \citenamefont {Peik}}]{huntemann_single-ion_2016}%
  \BibitemOpen
  \bibfield  {author} {\bibinfo {author} {\bibfnamefont {N.}~\bibnamefont
  {Huntemann}}, \bibinfo {author} {\bibfnamefont {C.}~\bibnamefont {Sanner}},
  \bibinfo {author} {\bibfnamefont {B.}~\bibnamefont {Lipphardt}}, \bibinfo
  {author} {\bibfnamefont {C.}~\bibnamefont {Tamm}}, \ and\ \bibinfo {author}
  {\bibfnamefont {E.}~\bibnamefont {Peik}},\ }\href {\doibase
  10.1103/PhysRevLett.116.063001} {\bibfield  {journal} {\bibinfo  {journal}
  {Phys. Rev. Lett.}\ }\textbf {\bibinfo {volume} {116}},\ \bibinfo {pages}
  {063001} (\bibinfo {year} {2016})}\BibitemShut {NoStop}%
\bibitem [{\citenamefont {Nicholson}\ \emph {et~al.}(2015)\citenamefont
  {Nicholson}, \citenamefont {Campbell}, \citenamefont {Hutson}, \citenamefont
  {Marti}, \citenamefont {Bloom}, \citenamefont {McNally}, \citenamefont
  {Zhang}, \citenamefont {Barrett}, \citenamefont {Safronova}, \citenamefont
  {Strouse}, \citenamefont {Tew},\ and\ \citenamefont
  {Ye}}]{nicholson_systematic_2015}%
  \BibitemOpen
  \bibfield  {author} {\bibinfo {author} {\bibfnamefont {T.~L.}\ \bibnamefont
  {Nicholson}}, \bibinfo {author} {\bibfnamefont {S.~L.}\ \bibnamefont
  {Campbell}}, \bibinfo {author} {\bibfnamefont {R.~B.}\ \bibnamefont
  {Hutson}}, \bibinfo {author} {\bibfnamefont {G.~E.}\ \bibnamefont {Marti}},
  \bibinfo {author} {\bibfnamefont {B.~J.}\ \bibnamefont {Bloom}}, \bibinfo
  {author} {\bibfnamefont {R.~L.}\ \bibnamefont {McNally}}, \bibinfo {author}
  {\bibfnamefont {W.}~\bibnamefont {Zhang}}, \bibinfo {author} {\bibfnamefont
  {M.~D.}\ \bibnamefont {Barrett}}, \bibinfo {author} {\bibfnamefont {M.~S.}\
  \bibnamefont {Safronova}}, \bibinfo {author} {\bibfnamefont {G.~F.}\
  \bibnamefont {Strouse}}, \bibinfo {author} {\bibfnamefont {W.~L.}\
  \bibnamefont {Tew}}, \ and\ \bibinfo {author} {\bibfnamefont
  {J.}~\bibnamefont {Ye}},\ }\href {\doibase 10.1038/ncomms7896} {\bibfield
  {journal} {\bibinfo  {journal} {Nat. Commun.}\ }\textbf {\bibinfo {volume}
  {6}},\ \bibinfo {pages} {6896} (\bibinfo {year} {2015})}\BibitemShut
  {NoStop}%
\bibitem [{\citenamefont {Bjerhammar}(1985)}]{bjerhammar_relativistic_1985}%
  \BibitemOpen
  \bibfield  {author} {\bibinfo {author} {\bibfnamefont {A.}~\bibnamefont
  {Bjerhammar}},\ }\href {\doibase 10.1007/BF02520327} {\bibfield  {journal}
  {\bibinfo  {journal} {Bulletin Géodésique}\ }\textbf {\bibinfo {volume}
  {59}},\ \bibinfo {pages} {207} (\bibinfo {year} {1985})}\BibitemShut
  {NoStop}%
\bibitem [{\citenamefont {Koller}\ \emph {et~al.}(2017)\citenamefont {Koller},
  \citenamefont {Grotti}, \citenamefont {Vogt}, \citenamefont {Al-Masoudi},
  \citenamefont {D{\"o}rscher}, \citenamefont {H{\"a}fner}, \citenamefont
  {Sterr},\ and\ \citenamefont {Lisdat}}]{koller_transportable_2017}%
  \BibitemOpen
  \bibfield  {author} {\bibinfo {author} {\bibfnamefont {S.~B.}\ \bibnamefont
  {Koller}}, \bibinfo {author} {\bibfnamefont {J.}~\bibnamefont {Grotti}},
  \bibinfo {author} {\bibfnamefont {S.}~\bibnamefont {Vogt}}, \bibinfo {author}
  {\bibfnamefont {A.}~\bibnamefont {Al-Masoudi}}, \bibinfo {author}
  {\bibfnamefont {S.}~\bibnamefont {D{\"o}rscher}}, \bibinfo {author}
  {\bibfnamefont {S.}~\bibnamefont {H{\"a}fner}}, \bibinfo {author}
  {\bibfnamefont {U.}~\bibnamefont {Sterr}}, \ and\ \bibinfo {author}
  {\bibfnamefont {C.}~\bibnamefont {Lisdat}},\ }\href {\doibase
  10.1103/PhysRevLett.118.073601} {\bibfield  {journal} {\bibinfo  {journal}
  {Phys. Rev. Lett.}\ }\textbf {\bibinfo {volume} {118}},\ \bibinfo {pages}
  {073601} (\bibinfo {year} {2017})}\BibitemShut {NoStop}%
\bibitem [{\citenamefont {Grotti}\ \emph {et~al.}(2017)\citenamefont {Grotti},
  \citenamefont {Koller}, \citenamefont {Vogt}, \citenamefont {H{\"a}fner},
  \citenamefont {Sterr}, \citenamefont {Lisdat}, \citenamefont {Denker},
  \citenamefont {Voigt}, \citenamefont {Timmen}, \citenamefont {Rolland},
  \citenamefont {Baynes}, \citenamefont {Margolis}, \citenamefont {Zampaolo},
  \citenamefont {Thoumany}, \citenamefont {Pizzocaro}, \citenamefont {Rauf},
  \citenamefont {Bregolin}, \citenamefont {Tampellini}, \citenamefont
  {Barbieri}, \citenamefont {Zucco}, \citenamefont {Costanzo}, \citenamefont
  {Clivati}, \citenamefont {Levi},\ and\ \citenamefont
  {Calonico}}]{grotti_geodesy_2017}%
  \BibitemOpen
  \bibfield  {author} {\bibinfo {author} {\bibfnamefont {J.}~\bibnamefont
  {Grotti}}, \bibinfo {author} {\bibfnamefont {S.}~\bibnamefont {Koller}},
  \bibinfo {author} {\bibfnamefont {S.}~\bibnamefont {Vogt}}, \bibinfo {author}
  {\bibfnamefont {S.}~\bibnamefont {H{\"a}fner}}, \bibinfo {author}
  {\bibfnamefont {U.}~\bibnamefont {Sterr}}, \bibinfo {author} {\bibfnamefont
  {C.}~\bibnamefont {Lisdat}}, \bibinfo {author} {\bibfnamefont
  {H.}~\bibnamefont {Denker}}, \bibinfo {author} {\bibfnamefont
  {C.}~\bibnamefont {Voigt}}, \bibinfo {author} {\bibfnamefont
  {L.}~\bibnamefont {Timmen}}, \bibinfo {author} {\bibfnamefont
  {A.}~\bibnamefont {Rolland}}, \bibinfo {author} {\bibfnamefont {F.~N.}\
  \bibnamefont {Baynes}}, \bibinfo {author} {\bibfnamefont {H.~S.}\
  \bibnamefont {Margolis}}, \bibinfo {author} {\bibfnamefont {M.}~\bibnamefont
  {Zampaolo}}, \bibinfo {author} {\bibfnamefont {P.}~\bibnamefont {Thoumany}},
  \bibinfo {author} {\bibfnamefont {M.}~\bibnamefont {Pizzocaro}}, \bibinfo
  {author} {\bibfnamefont {B.}~\bibnamefont {Rauf}}, \bibinfo {author}
  {\bibfnamefont {F.}~\bibnamefont {Bregolin}}, \bibinfo {author}
  {\bibfnamefont {A.}~\bibnamefont {Tampellini}}, \bibinfo {author}
  {\bibfnamefont {P.}~\bibnamefont {Barbieri}}, \bibinfo {author}
  {\bibfnamefont {M.}~\bibnamefont {Zucco}}, \bibinfo {author} {\bibfnamefont
  {G.~A.}\ \bibnamefont {Costanzo}}, \bibinfo {author} {\bibfnamefont
  {C.}~\bibnamefont {Clivati}}, \bibinfo {author} {\bibfnamefont
  {F.}~\bibnamefont {Levi}}, \ and\ \bibinfo {author} {\bibfnamefont
  {D.}~\bibnamefont {Calonico}},\ }\href {http://arxiv.org/abs/1705.04089}
  {\bibfield  {journal} {\bibinfo  {journal} {arXiv:1705.04089 [physics]}\ }
  (\bibinfo {year} {2017})},\ \bibinfo {note} {arXiv: 1705.04089}\BibitemShut
  {NoStop}%
\bibitem [{\citenamefont {Cao}\ \emph {et~al.}(2016)\citenamefont {Cao},
  \citenamefont {Zhang}, \citenamefont {Shang}, \citenamefont {Cui},
  \citenamefont {Yuan}, \citenamefont {Chao}, \citenamefont {Wang},
  \citenamefont {Shu},\ and\ \citenamefont {Huang}}]{cao_transportable_2016}%
  \BibitemOpen
  \bibfield  {author} {\bibinfo {author} {\bibfnamefont {J.}~\bibnamefont
  {Cao}}, \bibinfo {author} {\bibfnamefont {P.}~\bibnamefont {Zhang}}, \bibinfo
  {author} {\bibfnamefont {J.}~\bibnamefont {Shang}}, \bibinfo {author}
  {\bibfnamefont {K.}~\bibnamefont {Cui}}, \bibinfo {author} {\bibfnamefont
  {J.}~\bibnamefont {Yuan}}, \bibinfo {author} {\bibfnamefont {S.}~\bibnamefont
  {Chao}}, \bibinfo {author} {\bibfnamefont {S.}~\bibnamefont {Wang}}, \bibinfo
  {author} {\bibfnamefont {H.}~\bibnamefont {Shu}}, \ and\ \bibinfo {author}
  {\bibfnamefont {X.}~\bibnamefont {Huang}},\ }\href
  {http://arxiv.org/abs/1607.03731} {\bibfield  {journal} {\bibinfo  {journal}
  {arXiv:1607.03731}\ } (\bibinfo {year} {2016})},\ \bibinfo {note} {arXiv:
  1607.03731}\BibitemShut {NoStop}%
\bibitem [{\citenamefont {M{\"u}ntinga}\ \emph {et~al.}(2013)\citenamefont
  {M{\"u}ntinga}, \citenamefont {Ahlers}, \citenamefont {Krutzik},
  \citenamefont {Wenzlawski}, \citenamefont {Arnold}, \citenamefont {Becker},
  \citenamefont {Bongs}, \citenamefont {Dittus}, \citenamefont {Duncker},
  \citenamefont {Gaaloul}, \citenamefont {Gherasim}, \citenamefont {Giese},
  \citenamefont {Grzeschik}, \citenamefont {H{\"a}nsch}, \citenamefont
  {Hellmig}, \citenamefont {Herr}, \citenamefont {Herrmann}, \citenamefont
  {Kajari}, \citenamefont {Kleinert}, \citenamefont {L{\"a}mmerzahl},
  \citenamefont {Lewoczko-Adamczyk}, \citenamefont {Malcolm}, \citenamefont
  {Meyer}, \citenamefont {Nolte}, \citenamefont {Peters}, \citenamefont {Popp},
  \citenamefont {Reichel}, \citenamefont {Roura}, \citenamefont {Rudolph},
  \citenamefont {Schiemangk}, \citenamefont {Schneider}, \citenamefont
  {Seidel}, \citenamefont {Sengstock}, \citenamefont {Tamma}, \citenamefont
  {Valenzuela}, \citenamefont {Vogel}, \citenamefont {Walser}, \citenamefont
  {Wendrich}, \citenamefont {Windpassinger}, \citenamefont {Zeller},
  \citenamefont {van Zoest}, \citenamefont {Ertmer}, \citenamefont {Schleich},\
  and\ \citenamefont {Rasel}}]{muntinga_interferometry_2013}%
  \BibitemOpen
  \bibfield  {author} {\bibinfo {author} {\bibfnamefont {H.}~\bibnamefont
  {M{\"u}ntinga}}, \bibinfo {author} {\bibfnamefont {H.}~\bibnamefont
  {Ahlers}}, \bibinfo {author} {\bibfnamefont {M.}~\bibnamefont {Krutzik}},
  \bibinfo {author} {\bibfnamefont {A.}~\bibnamefont {Wenzlawski}}, \bibinfo
  {author} {\bibfnamefont {S.}~\bibnamefont {Arnold}}, \bibinfo {author}
  {\bibfnamefont {D.}~\bibnamefont {Becker}}, \bibinfo {author} {\bibfnamefont
  {K.}~\bibnamefont {Bongs}}, \bibinfo {author} {\bibfnamefont
  {H.}~\bibnamefont {Dittus}}, \bibinfo {author} {\bibfnamefont
  {H.}~\bibnamefont {Duncker}}, \bibinfo {author} {\bibfnamefont
  {N.}~\bibnamefont {Gaaloul}}, \bibinfo {author} {\bibfnamefont
  {C.}~\bibnamefont {Gherasim}}, \bibinfo {author} {\bibfnamefont
  {E.}~\bibnamefont {Giese}}, \bibinfo {author} {\bibfnamefont
  {C.}~\bibnamefont {Grzeschik}}, \bibinfo {author} {\bibfnamefont {T.~W.}\
  \bibnamefont {H{\"a}nsch}}, \bibinfo {author} {\bibfnamefont
  {O.}~\bibnamefont {Hellmig}}, \bibinfo {author} {\bibfnamefont
  {W.}~\bibnamefont {Herr}}, \bibinfo {author} {\bibfnamefont {S.}~\bibnamefont
  {Herrmann}}, \bibinfo {author} {\bibfnamefont {E.}~\bibnamefont {Kajari}},
  \bibinfo {author} {\bibfnamefont {S.}~\bibnamefont {Kleinert}}, \bibinfo
  {author} {\bibfnamefont {C.}~\bibnamefont {L{\"a}mmerzahl}}, \bibinfo
  {author} {\bibfnamefont {W.}~\bibnamefont {Lewoczko-Adamczyk}}, \bibinfo
  {author} {\bibfnamefont {J.}~\bibnamefont {Malcolm}}, \bibinfo {author}
  {\bibfnamefont {N.}~\bibnamefont {Meyer}}, \bibinfo {author} {\bibfnamefont
  {R.}~\bibnamefont {Nolte}}, \bibinfo {author} {\bibfnamefont
  {A.}~\bibnamefont {Peters}}, \bibinfo {author} {\bibfnamefont
  {M.}~\bibnamefont {Popp}}, \bibinfo {author} {\bibfnamefont {J.}~\bibnamefont
  {Reichel}}, \bibinfo {author} {\bibfnamefont {A.}~\bibnamefont {Roura}},
  \bibinfo {author} {\bibfnamefont {J.}~\bibnamefont {Rudolph}}, \bibinfo
  {author} {\bibfnamefont {M.}~\bibnamefont {Schiemangk}}, \bibinfo {author}
  {\bibfnamefont {M.}~\bibnamefont {Schneider}}, \bibinfo {author}
  {\bibfnamefont {S.~T.}\ \bibnamefont {Seidel}}, \bibinfo {author}
  {\bibfnamefont {K.}~\bibnamefont {Sengstock}}, \bibinfo {author}
  {\bibfnamefont {V.}~\bibnamefont {Tamma}}, \bibinfo {author} {\bibfnamefont
  {T.}~\bibnamefont {Valenzuela}}, \bibinfo {author} {\bibfnamefont
  {A.}~\bibnamefont {Vogel}}, \bibinfo {author} {\bibfnamefont
  {R.}~\bibnamefont {Walser}}, \bibinfo {author} {\bibfnamefont
  {T.}~\bibnamefont {Wendrich}}, \bibinfo {author} {\bibfnamefont
  {P.}~\bibnamefont {Windpassinger}}, \bibinfo {author} {\bibfnamefont
  {W.}~\bibnamefont {Zeller}}, \bibinfo {author} {\bibfnamefont
  {T.}~\bibnamefont {van Zoest}}, \bibinfo {author} {\bibfnamefont
  {W.}~\bibnamefont {Ertmer}}, \bibinfo {author} {\bibfnamefont {W.~P.}\
  \bibnamefont {Schleich}}, \ and\ \bibinfo {author} {\bibfnamefont {E.~M.}\
  \bibnamefont {Rasel}},\ }\href {\doibase 10.1103/PhysRevLett.110.093602}
  {\bibfield  {journal} {\bibinfo  {journal} {Phys. Rev. Lett.}\ }\textbf
  {\bibinfo {volume} {110}},\ \bibinfo {pages} {093602} (\bibinfo {year}
  {2013})}\BibitemShut {NoStop}%
\bibitem [{\citenamefont {Origlia}\ \emph {et~al.}(2016)\citenamefont
  {Origlia}, \citenamefont {Schiller}, \citenamefont {Pramod}, \citenamefont
  {Smith}, \citenamefont {Singh}, \citenamefont {He}, \citenamefont {Viswam},
  \citenamefont {Świerad}, \citenamefont {Hughes}, \citenamefont {Bongs},
  \citenamefont {Sterr}, \citenamefont {Lisdat}, \citenamefont {Vogt},
  \citenamefont {Bize}, \citenamefont {Lodewyck}, \citenamefont {Targat},
  \citenamefont {Holleville}, \citenamefont {Venon}, \citenamefont {Gill},
  \citenamefont {Barwood}, \citenamefont {Hill}, \citenamefont {Ovchinnikov},
  \citenamefont {Kulosa}, \citenamefont {Ertmer}, \citenamefont {Rasel},
  \citenamefont {Stuhler}, \citenamefont {Kaenders},\ and\ \citenamefont
  {contributors}}]{origlia_development_2016}%
  \BibitemOpen
  \bibfield  {author} {\bibinfo {author} {\bibfnamefont {S.}~\bibnamefont
  {Origlia}}, \bibinfo {author} {\bibfnamefont {S.}~\bibnamefont {Schiller}},
  \bibinfo {author} {\bibfnamefont {M.~S.}\ \bibnamefont {Pramod}}, \bibinfo
  {author} {\bibfnamefont {L.}~\bibnamefont {Smith}}, \bibinfo {author}
  {\bibfnamefont {Y.}~\bibnamefont {Singh}}, \bibinfo {author} {\bibfnamefont
  {W.}~\bibnamefont {He}}, \bibinfo {author} {\bibfnamefont {S.}~\bibnamefont
  {Viswam}}, \bibinfo {author} {\bibfnamefont {D.}~\bibnamefont {Świerad}},
  \bibinfo {author} {\bibfnamefont {J.}~\bibnamefont {Hughes}}, \bibinfo
  {author} {\bibfnamefont {K.}~\bibnamefont {Bongs}}, \bibinfo {author}
  {\bibfnamefont {U.}~\bibnamefont {Sterr}}, \bibinfo {author} {\bibfnamefont
  {C.}~\bibnamefont {Lisdat}}, \bibinfo {author} {\bibfnamefont
  {S.}~\bibnamefont {Vogt}}, \bibinfo {author} {\bibfnamefont {S.}~\bibnamefont
  {Bize}}, \bibinfo {author} {\bibfnamefont {J.}~\bibnamefont {Lodewyck}},
  \bibinfo {author} {\bibfnamefont {R.~L.}\ \bibnamefont {Targat}}, \bibinfo
  {author} {\bibfnamefont {D.}~\bibnamefont {Holleville}}, \bibinfo {author}
  {\bibfnamefont {B.}~\bibnamefont {Venon}}, \bibinfo {author} {\bibfnamefont
  {P.}~\bibnamefont {Gill}}, \bibinfo {author} {\bibfnamefont {G.}~\bibnamefont
  {Barwood}}, \bibinfo {author} {\bibfnamefont {I.~R.}\ \bibnamefont {Hill}},
  \bibinfo {author} {\bibfnamefont {Y.}~\bibnamefont {Ovchinnikov}}, \bibinfo
  {author} {\bibfnamefont {A.}~\bibnamefont {Kulosa}}, \bibinfo {author}
  {\bibfnamefont {W.}~\bibnamefont {Ertmer}}, \bibinfo {author} {\bibfnamefont
  {E.-M.}\ \bibnamefont {Rasel}}, \bibinfo {author} {\bibfnamefont
  {J.}~\bibnamefont {Stuhler}}, \bibinfo {author} {\bibfnamefont
  {W.}~\bibnamefont {Kaenders}}, \ and\ \bibinfo {author} {\bibfnamefont
  {t.~S.~c.}\ \bibnamefont {contributors}},\ }\href {\doibase
  10.1117/12.2229473} {\bibfield  {journal} {\bibinfo  {journal} {Proc.SPIE}\
  }\textbf {\bibinfo {volume} {9900}},\ \bibinfo {pages} {9900 } (\bibinfo
  {year} {2016})}\BibitemShut {NoStop}%
\bibitem [{\citenamefont {Yin}\ \emph {et~al.}(2017)\citenamefont {Yin},
  \citenamefont {Cao}, \citenamefont {Li}, \citenamefont {Liao}, \citenamefont
  {Zhang}, \citenamefont {Ren}, \citenamefont {Cai}, \citenamefont {Liu},
  \citenamefont {Li}, \citenamefont {Dai}, \citenamefont {Li}, \citenamefont
  {Lu}, \citenamefont {Gong}, \citenamefont {Xu}, \citenamefont {Li},
  \citenamefont {Li}, \citenamefont {Yin}, \citenamefont {Jiang}, \citenamefont
  {Li}, \citenamefont {Jia}, \citenamefont {Ren}, \citenamefont {He},
  \citenamefont {Zhou}, \citenamefont {Zhang}, \citenamefont {Wang},
  \citenamefont {Chang}, \citenamefont {Zhu}, \citenamefont {Liu},
  \citenamefont {Chen}, \citenamefont {Lu}, \citenamefont {Shu}, \citenamefont
  {Peng}, \citenamefont {Wang},\ and\ \citenamefont
  {Pan}}]{yin_satellite-based_2017}%
  \BibitemOpen
  \bibfield  {author} {\bibinfo {author} {\bibfnamefont {J.}~\bibnamefont
  {Yin}}, \bibinfo {author} {\bibfnamefont {Y.}~\bibnamefont {Cao}}, \bibinfo
  {author} {\bibfnamefont {Y.-H.}\ \bibnamefont {Li}}, \bibinfo {author}
  {\bibfnamefont {S.-K.}\ \bibnamefont {Liao}}, \bibinfo {author}
  {\bibfnamefont {L.}~\bibnamefont {Zhang}}, \bibinfo {author} {\bibfnamefont
  {J.-G.}\ \bibnamefont {Ren}}, \bibinfo {author} {\bibfnamefont {W.-Q.}\
  \bibnamefont {Cai}}, \bibinfo {author} {\bibfnamefont {W.-Y.}\ \bibnamefont
  {Liu}}, \bibinfo {author} {\bibfnamefont {B.}~\bibnamefont {Li}}, \bibinfo
  {author} {\bibfnamefont {H.}~\bibnamefont {Dai}}, \bibinfo {author}
  {\bibfnamefont {G.-B.}\ \bibnamefont {Li}}, \bibinfo {author} {\bibfnamefont
  {Q.-M.}\ \bibnamefont {Lu}}, \bibinfo {author} {\bibfnamefont {Y.-H.}\
  \bibnamefont {Gong}}, \bibinfo {author} {\bibfnamefont {Y.}~\bibnamefont
  {Xu}}, \bibinfo {author} {\bibfnamefont {S.-L.}\ \bibnamefont {Li}}, \bibinfo
  {author} {\bibfnamefont {F.-Z.}\ \bibnamefont {Li}}, \bibinfo {author}
  {\bibfnamefont {Y.-Y.}\ \bibnamefont {Yin}}, \bibinfo {author} {\bibfnamefont
  {Z.-Q.}\ \bibnamefont {Jiang}}, \bibinfo {author} {\bibfnamefont
  {M.}~\bibnamefont {Li}}, \bibinfo {author} {\bibfnamefont {J.-J.}\
  \bibnamefont {Jia}}, \bibinfo {author} {\bibfnamefont {G.}~\bibnamefont
  {Ren}}, \bibinfo {author} {\bibfnamefont {D.}~\bibnamefont {He}}, \bibinfo
  {author} {\bibfnamefont {Y.-L.}\ \bibnamefont {Zhou}}, \bibinfo {author}
  {\bibfnamefont {X.-X.}\ \bibnamefont {Zhang}}, \bibinfo {author}
  {\bibfnamefont {N.}~\bibnamefont {Wang}}, \bibinfo {author} {\bibfnamefont
  {X.}~\bibnamefont {Chang}}, \bibinfo {author} {\bibfnamefont {Z.-C.}\
  \bibnamefont {Zhu}}, \bibinfo {author} {\bibfnamefont {N.-L.}\ \bibnamefont
  {Liu}}, \bibinfo {author} {\bibfnamefont {Y.-A.}\ \bibnamefont {Chen}},
  \bibinfo {author} {\bibfnamefont {C.-Y.}\ \bibnamefont {Lu}}, \bibinfo
  {author} {\bibfnamefont {R.}~\bibnamefont {Shu}}, \bibinfo {author}
  {\bibfnamefont {C.-Z.}\ \bibnamefont {Peng}}, \bibinfo {author}
  {\bibfnamefont {J.-Y.}\ \bibnamefont {Wang}}, \ and\ \bibinfo {author}
  {\bibfnamefont {J.-W.}\ \bibnamefont {Pan}},\ }\href {\doibase
  10.1126/science.aan3211} {\bibfield  {journal} {\bibinfo  {journal}
  {Science}\ }\textbf {\bibinfo {volume} {356}},\ \bibinfo {pages} {1140}
  (\bibinfo {year} {2017})}\BibitemShut {NoStop}%
\bibitem [{\citenamefont {Luvsandamdin}\ \emph {et~al.}(2014)\citenamefont
  {Luvsandamdin}, \citenamefont {K{\"u}rbis}, \citenamefont {Schiemangk},
  \citenamefont {Sahm}, \citenamefont {Wicht}, \citenamefont {Peters},
  \citenamefont {Erbert},\ and\ \citenamefont
  {Tr{\"a}nkle}}]{luvsandamdin_micro-integrated_2014}%
  \BibitemOpen
  \bibfield  {author} {\bibinfo {author} {\bibfnamefont {E.}~\bibnamefont
  {Luvsandamdin}}, \bibinfo {author} {\bibfnamefont {C.}~\bibnamefont
  {K{\"u}rbis}}, \bibinfo {author} {\bibfnamefont {M.}~\bibnamefont
  {Schiemangk}}, \bibinfo {author} {\bibfnamefont {A.}~\bibnamefont {Sahm}},
  \bibinfo {author} {\bibfnamefont {A.}~\bibnamefont {Wicht}}, \bibinfo
  {author} {\bibfnamefont {A.}~\bibnamefont {Peters}}, \bibinfo {author}
  {\bibfnamefont {G.}~\bibnamefont {Erbert}}, \ and\ \bibinfo {author}
  {\bibfnamefont {G.}~\bibnamefont {Tr{\"a}nkle}},\ }\href {\doibase
  10.1364/OE.22.007790} {\bibfield  {journal} {\bibinfo  {journal} {Opt.
  Express}\ }\textbf {\bibinfo {volume} {22}},\ \bibinfo {pages} {7790}
  (\bibinfo {year} {2014})}\BibitemShut {NoStop}%
\bibitem [{\citenamefont {Kohfeldt}\ \emph {et~al.}(2016)\citenamefont
  {Kohfeldt}, \citenamefont {K{\"u}rbis}, \citenamefont {Luvsandamdin},
  \citenamefont {Schiemangk}, \citenamefont {Wicht}, \citenamefont {Peters},
  \citenamefont {Erbert},\ and\ \citenamefont
  {Tr{\"a}nkle}}]{kohfeldt_compact_2016}%
  \BibitemOpen
  \bibfield  {author} {\bibinfo {author} {\bibfnamefont {A.}~\bibnamefont
  {Kohfeldt}}, \bibinfo {author} {\bibfnamefont {C.}~\bibnamefont
  {K{\"u}rbis}}, \bibinfo {author} {\bibfnamefont {E.}~\bibnamefont
  {Luvsandamdin}}, \bibinfo {author} {\bibfnamefont {M.}~\bibnamefont
  {Schiemangk}}, \bibinfo {author} {\bibfnamefont {A.}~\bibnamefont {Wicht}},
  \bibinfo {author} {\bibfnamefont {A.}~\bibnamefont {Peters}}, \bibinfo
  {author} {\bibfnamefont {G.}~\bibnamefont {Erbert}}, \ and\ \bibinfo {author}
  {\bibfnamefont {G.}~\bibnamefont {Tr{\"a}nkle}},\ }\href {\doibase
  10.1117/12.2231537} {\bibfield  {journal} {\bibinfo  {journal} {Proc. SPIE}\
  }\textbf {\bibinfo {volume} {9900}},\ \bibinfo {pages} {9900 } (\bibinfo
  {year} {2016})}\BibitemShut {NoStop}%
\bibitem [{\citenamefont {Wakui}, \citenamefont {Hayasaka},\ and\ \citenamefont
  {Ido}(2014)}]{wakui_generation_2014}%
  \BibitemOpen
  \bibfield  {author} {\bibinfo {author} {\bibfnamefont {K.}~\bibnamefont
  {Wakui}}, \bibinfo {author} {\bibfnamefont {K.}~\bibnamefont {Hayasaka}}, \
  and\ \bibinfo {author} {\bibfnamefont {T.}~\bibnamefont {Ido}},\ }\href
  {\doibase 10.1007/s00340-014-5914-y} {\bibfield  {journal} {\bibinfo
  {journal} {Appl. Phys. B.}\ }\textbf {\bibinfo {volume} {117}},\ \bibinfo
  {pages} {957} (\bibinfo {year} {2014})}\BibitemShut {NoStop}%
\bibitem [{\citenamefont {Hu}\ \emph {et~al.}(2013)\citenamefont {Hu},
  \citenamefont {Zhang}, \citenamefont {Liu}, \citenamefont {Liu},
  \citenamefont {Xu},\ and\ \citenamefont {Feng}}]{hu_high_2013}%
  \BibitemOpen
  \bibfield  {author} {\bibinfo {author} {\bibfnamefont {J.}~\bibnamefont
  {Hu}}, \bibinfo {author} {\bibfnamefont {L.}~\bibnamefont {Zhang}}, \bibinfo
  {author} {\bibfnamefont {H.}~\bibnamefont {Liu}}, \bibinfo {author}
  {\bibfnamefont {K.}~\bibnamefont {Liu}}, \bibinfo {author} {\bibfnamefont
  {Z.}~\bibnamefont {Xu}}, \ and\ \bibinfo {author} {\bibfnamefont
  {Y.}~\bibnamefont {Feng}},\ }\href {\doibase 10.1364/OE.21.030958} {\bibfield
   {journal} {\bibinfo  {journal} {Opt. Express}\ }\textbf {\bibinfo {volume}
  {21}},\ \bibinfo {pages} {30958} (\bibinfo {year} {2013})}\BibitemShut
  {NoStop}%
\bibitem [{\citenamefont {Sherstov}\ \emph {et~al.}(2010)\citenamefont
  {Sherstov}, \citenamefont {Okhapkin}, \citenamefont {Lipphardt},
  \citenamefont {Tamm},\ and\ \citenamefont
  {Peik}}]{sherstov_diode-laser_2010}%
  \BibitemOpen
  \bibfield  {author} {\bibinfo {author} {\bibfnamefont {I.}~\bibnamefont
  {Sherstov}}, \bibinfo {author} {\bibfnamefont {M.}~\bibnamefont {Okhapkin}},
  \bibinfo {author} {\bibfnamefont {B.}~\bibnamefont {Lipphardt}}, \bibinfo
  {author} {\bibfnamefont {C.}~\bibnamefont {Tamm}}, \ and\ \bibinfo {author}
  {\bibfnamefont {E.}~\bibnamefont {Peik}},\ }\href {\doibase
  10.1103/PhysRevA.81.021805} {\bibfield  {journal} {\bibinfo  {journal} {Phys.
  Rev. A}\ }\textbf {\bibinfo {volume} {81}},\ \bibinfo {pages} {021805}
  (\bibinfo {year} {2010})}\BibitemShut {NoStop}%
\bibitem [{\citenamefont {Scheid}\ \emph {et~al.}(2007)\citenamefont {Scheid},
  \citenamefont {Markert}, \citenamefont {Walz}, \citenamefont {Wang},
  \citenamefont {Kirchner},\ and\ \citenamefont
  {H{\"a}nsch}}]{scheid_750_2007}%
  \BibitemOpen
  \bibfield  {author} {\bibinfo {author} {\bibfnamefont {M.}~\bibnamefont
  {Scheid}}, \bibinfo {author} {\bibfnamefont {F.}~\bibnamefont {Markert}},
  \bibinfo {author} {\bibfnamefont {J.}~\bibnamefont {Walz}}, \bibinfo {author}
  {\bibfnamefont {J.}~\bibnamefont {Wang}}, \bibinfo {author} {\bibfnamefont
  {M.}~\bibnamefont {Kirchner}}, \ and\ \bibinfo {author} {\bibfnamefont
  {T.~W.}\ \bibnamefont {H{\"a}nsch}},\ }\href
  {http://www.opticsinfobase.org/ol/fulltext.cfm?uri=ol-32-8-955} {\bibfield
  {journal} {\bibinfo  {journal} {Opt. Lett.}\ }\textbf {\bibinfo {volume}
  {32}},\ \bibinfo {pages} {955} (\bibinfo {year} {2007})}\BibitemShut
  {NoStop}%
\bibitem [{\citenamefont {Wen}\ \emph {et~al.}(2014)\citenamefont {Wen},
  \citenamefont {Han}, \citenamefont {Bai}, \citenamefont {He}, \citenamefont
  {Wang}, \citenamefont {Yang},\ and\ \citenamefont
  {Wang}}]{wen_cavity-enhanced_2014}%
  \BibitemOpen
  \bibfield  {author} {\bibinfo {author} {\bibfnamefont {X.}~\bibnamefont
  {Wen}}, \bibinfo {author} {\bibfnamefont {Y.}~\bibnamefont {Han}}, \bibinfo
  {author} {\bibfnamefont {J.}~\bibnamefont {Bai}}, \bibinfo {author}
  {\bibfnamefont {J.}~\bibnamefont {He}}, \bibinfo {author} {\bibfnamefont
  {Y.}~\bibnamefont {Wang}}, \bibinfo {author} {\bibfnamefont {B.}~\bibnamefont
  {Yang}}, \ and\ \bibinfo {author} {\bibfnamefont {J.}~\bibnamefont {Wang}},\
  }\href {\doibase 10.1364/OE.22.032293} {\bibfield  {journal} {\bibinfo
  {journal} {Opt. Express}\ }\textbf {\bibinfo {volume} {22}},\ \bibinfo
  {pages} {32293} (\bibinfo {year} {2014})}\BibitemShut {NoStop}%
\bibitem [{\citenamefont {Eismann}\ \emph {et~al.}(2012)\citenamefont
  {Eismann}, \citenamefont {Gerbier}, \citenamefont {Canalias}, \citenamefont
  {Zukauskas}, \citenamefont {Trénec}, \citenamefont {Vigué}, \citenamefont
  {Chevy},\ and\ \citenamefont {Salomon}}]{eismann_all-solid-state_2012}%
  \BibitemOpen
  \bibfield  {author} {\bibinfo {author} {\bibfnamefont {U.}~\bibnamefont
  {Eismann}}, \bibinfo {author} {\bibfnamefont {F.}~\bibnamefont {Gerbier}},
  \bibinfo {author} {\bibfnamefont {C.}~\bibnamefont {Canalias}}, \bibinfo
  {author} {\bibfnamefont {A.}~\bibnamefont {Zukauskas}}, \bibinfo {author}
  {\bibfnamefont {G.}~\bibnamefont {Trénec}}, \bibinfo {author} {\bibfnamefont
  {J.}~\bibnamefont {Vigué}}, \bibinfo {author} {\bibfnamefont
  {F.}~\bibnamefont {Chevy}}, \ and\ \bibinfo {author} {\bibfnamefont
  {C.}~\bibnamefont {Salomon}},\ }\href {\doibase 10.1007/s00340-011-4693-y}
  {\bibfield  {journal} {\bibinfo  {journal} {Appl. Phys. B.}\ }\textbf
  {\bibinfo {volume} {106}},\ \bibinfo {pages} {25} (\bibinfo {year}
  {2012})}\BibitemShut {NoStop}%
\bibitem [{\citenamefont {Wilson}\ \emph {et~al.}(2011)\citenamefont {Wilson},
  \citenamefont {Ospelkaus}, \citenamefont {VanDevender}, \citenamefont
  {Mlynek}, \citenamefont {Brown}, \citenamefont {Leibfried},\ and\
  \citenamefont {Wineland}}]{wilson_750-mw_2011}%
  \BibitemOpen
  \bibfield  {author} {\bibinfo {author} {\bibfnamefont {A.~C.}\ \bibnamefont
  {Wilson}}, \bibinfo {author} {\bibfnamefont {C.}~\bibnamefont {Ospelkaus}},
  \bibinfo {author} {\bibfnamefont {A.~P.}\ \bibnamefont {VanDevender}},
  \bibinfo {author} {\bibfnamefont {J.~A.}\ \bibnamefont {Mlynek}}, \bibinfo
  {author} {\bibfnamefont {K.~R.}\ \bibnamefont {Brown}}, \bibinfo {author}
  {\bibfnamefont {D.}~\bibnamefont {Leibfried}}, \ and\ \bibinfo {author}
  {\bibfnamefont {D.~J.}\ \bibnamefont {Wineland}},\ }\href {\doibase
  10.1007/s00340-011-4771-1} {\bibfield  {journal} {\bibinfo  {journal} {Appl.
  Phys. B.}\ }\textbf {\bibinfo {volume} {105}},\ \bibinfo {pages} {741}
  (\bibinfo {year} {2011})}\BibitemShut {NoStop}%
\bibitem [{\citenamefont {Vasilyev}\ \emph {et~al.}(2011)\citenamefont
  {Vasilyev}, \citenamefont {Nevsky}, \citenamefont {Ernsting}, \citenamefont
  {Hansen}, \citenamefont {Shen},\ and\ \citenamefont
  {Schiller}}]{vasilyev_compact_2011}%
  \BibitemOpen
  \bibfield  {author} {\bibinfo {author} {\bibfnamefont {S.}~\bibnamefont
  {Vasilyev}}, \bibinfo {author} {\bibfnamefont {A.}~\bibnamefont {Nevsky}},
  \bibinfo {author} {\bibfnamefont {I.}~\bibnamefont {Ernsting}}, \bibinfo
  {author} {\bibfnamefont {M.}~\bibnamefont {Hansen}}, \bibinfo {author}
  {\bibfnamefont {J.}~\bibnamefont {Shen}}, \ and\ \bibinfo {author}
  {\bibfnamefont {S.}~\bibnamefont {Schiller}},\ }\href {\doibase
  10.1007/s00340-011-4435-1} {\bibfield  {journal} {\bibinfo  {journal} {Appl.
  Phys. B.}\ }\textbf {\bibinfo {volume} {103}},\ \bibinfo {pages} {27}
  (\bibinfo {year} {2011})}\BibitemShut {NoStop}%
\bibitem [{\citenamefont {Carollo}\ \emph {et~al.}(2017)\citenamefont
  {Carollo}, \citenamefont {Lane}, \citenamefont {Kleiner}, \citenamefont
  {Kyaw}, \citenamefont {Teng}, \citenamefont {Ou}, \citenamefont {Qiao},\ and\
  \citenamefont {Hanneke}}]{carollo_third-harmonic-generation_2017}%
  \BibitemOpen
  \bibfield  {author} {\bibinfo {author} {\bibfnamefont {R.~A.}\ \bibnamefont
  {Carollo}}, \bibinfo {author} {\bibfnamefont {D.~A.}\ \bibnamefont {Lane}},
  \bibinfo {author} {\bibfnamefont {E.~K.}\ \bibnamefont {Kleiner}}, \bibinfo
  {author} {\bibfnamefont {P.~A.}\ \bibnamefont {Kyaw}}, \bibinfo {author}
  {\bibfnamefont {C.~C.}\ \bibnamefont {Teng}}, \bibinfo {author}
  {\bibfnamefont {C.~Y.}\ \bibnamefont {Ou}}, \bibinfo {author} {\bibfnamefont
  {S.}~\bibnamefont {Qiao}}, \ and\ \bibinfo {author} {\bibfnamefont
  {D.}~\bibnamefont {Hanneke}},\ }\href {\doibase 10.1364/OE.25.007220}
  {\bibfield  {journal} {\bibinfo  {journal} {Opt. Express}\ }\textbf {\bibinfo
  {volume} {25}},\ \bibinfo {pages} {7220} (\bibinfo {year}
  {2017})}\BibitemShut {NoStop}%
\bibitem [{\citenamefont {Franken}\ \emph {et~al.}(1961)\citenamefont
  {Franken}, \citenamefont {Hill}, \citenamefont {Peters},\ and\ \citenamefont
  {Weinreich}}]{Franken_Optical_Harmonics_1961}%
  \BibitemOpen
  \bibfield  {author} {\bibinfo {author} {\bibfnamefont {P.~A.}\ \bibnamefont
  {Franken}}, \bibinfo {author} {\bibfnamefont {A.~E.}\ \bibnamefont {Hill}},
  \bibinfo {author} {\bibfnamefont {C.~W.}\ \bibnamefont {Peters}}, \ and\
  \bibinfo {author} {\bibfnamefont {G.}~\bibnamefont {Weinreich}},\ }\href
  {\doibase 10.1103/PhysRevLett.7.118} {\bibfield  {journal} {\bibinfo
  {journal} {Phys. Rev. Lett.}\ }\textbf {\bibinfo {volume} {7}},\ \bibinfo
  {pages} {118} (\bibinfo {year} {1961})}\BibitemShut {NoStop}%
\bibitem [{\citenamefont {Kane}\ and\ \citenamefont
  {Byer}(1985)}]{kane_monolithic_1985}%
  \BibitemOpen
  \bibfield  {author} {\bibinfo {author} {\bibfnamefont {T.~J.}\ \bibnamefont
  {Kane}}\ and\ \bibinfo {author} {\bibfnamefont {R.~L.}\ \bibnamefont
  {Byer}},\ }\href {\doibase 10.1364/OL.10.000065} {\bibfield  {journal}
  {\bibinfo  {journal} {Opt. Lett.}\ }\textbf {\bibinfo {volume} {10}},\
  \bibinfo {pages} {65} (\bibinfo {year} {1985})}\BibitemShut {NoStop}%
\bibitem [{\citenamefont {Kozlovsky}, \citenamefont {Nabors},\ and\
  \citenamefont {Byer}(1988)}]{kozlovsky_efficient_1988}%
  \BibitemOpen
  \bibfield  {author} {\bibinfo {author} {\bibfnamefont {W.}~\bibnamefont
  {Kozlovsky}}, \bibinfo {author} {\bibfnamefont {C.~D.}\ \bibnamefont
  {Nabors}}, \ and\ \bibinfo {author} {\bibfnamefont {R.}~\bibnamefont
  {Byer}},\ }\href {\doibase 10.1109/3.211} {\bibfield  {journal} {\bibinfo
  {journal} {‎IEEE J. Quant. Electron.}\ }\textbf {\bibinfo {volume} {24}},\
  \bibinfo {pages} {913} (\bibinfo {year} {1988})}\BibitemShut {NoStop}%
\bibitem [{\citenamefont {Gerstenberger}, \citenamefont {Tye},\ and\
  \citenamefont {Wallace}(1991)}]{gerstenberger_efficient_1991}%
  \BibitemOpen
  \bibfield  {author} {\bibinfo {author} {\bibfnamefont {D.~C.}\ \bibnamefont
  {Gerstenberger}}, \bibinfo {author} {\bibfnamefont {G.~E.}\ \bibnamefont
  {Tye}}, \ and\ \bibinfo {author} {\bibfnamefont {R.~W.}\ \bibnamefont
  {Wallace}},\ }\href {\doibase 10.1364/OL.16.000992} {\bibfield  {journal}
  {\bibinfo  {journal} {Opt. Lett.}\ }\textbf {\bibinfo {volume} {16}},\
  \bibinfo {pages} {992} (\bibinfo {year} {1991})}\BibitemShut {NoStop}%
\bibitem [{\citenamefont {Corporation}(2017)}]{NTT}%
  \BibitemOpen
  \bibfield  {author} {\bibinfo {author} {\bibfnamefont {N.~E.}\ \bibnamefont
  {Corporation}},\ }\href@noop {} {\enquote {\bibinfo {title} {{ Wavelength
  Conversion Module}},}\ } (\bibinfo {year} {2017}),\ \bibinfo {note}
  {\url{http://www.ntt-electronics.com/en/products/photonics/conversion-module.html}}\BibitemShut
  {NoStop}%
\bibitem [{\citenamefont {Kozlovsky}\ \emph {et~al.}(1994)\citenamefont
  {Kozlovsky}, \citenamefont {Risk}, \citenamefont {Lenth}, \citenamefont
  {Kim}, \citenamefont {Bona}, \citenamefont {Jaeckel},\ and\ \citenamefont
  {Webb}}]{kozlovsky_blue_1994}%
  \BibitemOpen
  \bibfield  {author} {\bibinfo {author} {\bibfnamefont {W.~J.}\ \bibnamefont
  {Kozlovsky}}, \bibinfo {author} {\bibfnamefont {W.~P.}\ \bibnamefont {Risk}},
  \bibinfo {author} {\bibfnamefont {W.}~\bibnamefont {Lenth}}, \bibinfo
  {author} {\bibfnamefont {B.~G.}\ \bibnamefont {Kim}}, \bibinfo {author}
  {\bibfnamefont {G.~L.}\ \bibnamefont {Bona}}, \bibinfo {author}
  {\bibfnamefont {H.}~\bibnamefont {Jaeckel}}, \ and\ \bibinfo {author}
  {\bibfnamefont {D.~J.}\ \bibnamefont {Webb}},\ }\href {\doibase
  10.1063/1.112286} {\bibfield  {journal} {\bibinfo  {journal} {Applied Physics
  Letters}\ }\textbf {\bibinfo {volume} {65}},\ \bibinfo {pages} {525}
  (\bibinfo {year} {1994})}\BibitemShut {NoStop}%
\bibitem [{\citenamefont {Hemmerich}, \citenamefont {Zimmermann},\ and\
  \citenamefont {H{\"a}nsch}(1994)}]{hemmerich_compact_1994}%
  \BibitemOpen
  \bibfield  {author} {\bibinfo {author} {\bibfnamefont {A.}~\bibnamefont
  {Hemmerich}}, \bibinfo {author} {\bibfnamefont {C.}~\bibnamefont
  {Zimmermann}}, \ and\ \bibinfo {author} {\bibfnamefont {T.~W.}\ \bibnamefont
  {H{\"a}nsch}},\ }\href {\doibase 10.1364/AO.33.000988} {\bibfield  {journal}
  {\bibinfo  {journal} {Appl. Opt.}\ }\textbf {\bibinfo {volume} {33}},\
  \bibinfo {pages} {988} (\bibinfo {year} {1994})}\BibitemShut {NoStop}%
\bibitem [{\citenamefont {Skoczowsky}\ \emph {et~al.}(2010)\citenamefont
  {Skoczowsky}, \citenamefont {Jechow}, \citenamefont {Menzel}, \citenamefont
  {Paschke},\ and\ \citenamefont {Erbert}}]{skoczowsky_efficient_2010}%
  \BibitemOpen
  \bibfield  {author} {\bibinfo {author} {\bibfnamefont {D.}~\bibnamefont
  {Skoczowsky}}, \bibinfo {author} {\bibfnamefont {A.}~\bibnamefont {Jechow}},
  \bibinfo {author} {\bibfnamefont {R.}~\bibnamefont {Menzel}}, \bibinfo
  {author} {\bibfnamefont {K.}~\bibnamefont {Paschke}}, \ and\ \bibinfo
  {author} {\bibfnamefont {G.}~\bibnamefont {Erbert}},\ }\href {\doibase
  10.1364/OL.35.000232} {\bibfield  {journal} {\bibinfo  {journal} {Opt.
  Lett.}\ }\textbf {\bibinfo {volume} {35}},\ \bibinfo {pages} {232} (\bibinfo
  {year} {2010})}\BibitemShut {NoStop}%
\bibitem [{\citenamefont {Zimmermann}, \citenamefont {Kallenbach},\ and\
  \citenamefont {H\"ansch}(1990)}]{ZI1990}%
  \BibitemOpen
  \bibfield  {author} {\bibinfo {author} {\bibfnamefont {C.}~\bibnamefont
  {Zimmermann}}, \bibinfo {author} {\bibfnamefont {R.}~\bibnamefont
  {Kallenbach}}, \ and\ \bibinfo {author} {\bibfnamefont {T.~W.}\ \bibnamefont
  {H\"ansch}},\ }\href {\doibase 10.1103/PhysRevLett.65.571} {\bibfield
  {journal} {\bibinfo  {journal} {Phys. Rev. Lett.}\ }\textbf {\bibinfo
  {volume} {65}},\ \bibinfo {pages} {571} (\bibinfo {year} {1990})}\BibitemShut
  {NoStop}%
\bibitem [{\citenamefont {Huang}\ \emph {et~al.}(2014)\citenamefont {Huang},
  \citenamefont {Le~Jeannic}, \citenamefont {Ruaudel}, \citenamefont {Morin},\
  and\ \citenamefont {Laurat}}]{HU2014}%
  \BibitemOpen
  \bibfield  {author} {\bibinfo {author} {\bibfnamefont {K.}~\bibnamefont
  {Huang}}, \bibinfo {author} {\bibfnamefont {H.}~\bibnamefont {Le~Jeannic}},
  \bibinfo {author} {\bibfnamefont {J.}~\bibnamefont {Ruaudel}}, \bibinfo
  {author} {\bibfnamefont {O.}~\bibnamefont {Morin}}, \ and\ \bibinfo {author}
  {\bibfnamefont {J.}~\bibnamefont {Laurat}},\ }\href {\doibase
  10.1063/1.4903869} {\bibfield  {journal} {\bibinfo  {journal} {Review of
  Scientific Instruments}\ }\textbf {\bibinfo {volume} {85}},\ \bibinfo {pages}
  {123112} (\bibinfo {year} {2014})}\BibitemShut {NoStop}%
\bibitem [{\citenamefont {Dietrich}\ and\ \citenamefont
  {Blinov}(2009)}]{DB2009}%
  \BibitemOpen
  \bibfield  {author} {\bibinfo {author} {\bibfnamefont {M.~R.}\ \bibnamefont
  {Dietrich}}\ and\ \bibinfo {author} {\bibfnamefont {B.~B.}\ \bibnamefont
  {Blinov}},\ }\href {http://arxiv.org/abs/0905.2484} {\bibfield  {journal}
  {\bibinfo  {journal} {arXiv:0905.2484 [physics.atom-ph]}\ } (\bibinfo {year}
  {2009})}\BibitemShut {NoStop}%
\bibitem [{\citenamefont {Sparkes}\ \emph {et~al.}(2011)\citenamefont
  {Sparkes}, \citenamefont {Chrzanowski}, \citenamefont {Parrain},
  \citenamefont {Buchler}, \citenamefont {Lam},\ and\ \citenamefont
  {Symul}}]{sparkes_scalable_2011}%
  \BibitemOpen
  \bibfield  {author} {\bibinfo {author} {\bibfnamefont {B.~M.}\ \bibnamefont
  {Sparkes}}, \bibinfo {author} {\bibfnamefont {H.~M.}\ \bibnamefont
  {Chrzanowski}}, \bibinfo {author} {\bibfnamefont {D.~P.}\ \bibnamefont
  {Parrain}}, \bibinfo {author} {\bibfnamefont {B.~C.}\ \bibnamefont
  {Buchler}}, \bibinfo {author} {\bibfnamefont {P.~K.}\ \bibnamefont {Lam}}, \
  and\ \bibinfo {author} {\bibfnamefont {T.}~\bibnamefont {Symul}},\ }\href
  {http://arxiv.org/abs/1105.3795} {\bibfield  {journal} {\bibinfo  {journal}
  {arXiv:1105.3795 [physics]}\ } (\bibinfo {year} {2011})},\ \bibinfo {note}
  {arXiv: 1105.3795}\BibitemShut {NoStop}%
\bibitem [{\citenamefont {Leibrandt}\ and\ \citenamefont
  {Heidecker}(2015)}]{LH2015}%
  \BibitemOpen
  \bibfield  {author} {\bibinfo {author} {\bibfnamefont {D.~R.}\ \bibnamefont
  {Leibrandt}}\ and\ \bibinfo {author} {\bibfnamefont {J.}~\bibnamefont
  {Heidecker}},\ }\href {http://dx.doi.org/10.1063/1.4938282} {\bibfield
  {journal} {\bibinfo  {journal} {{Rev. Sci. Instrum.}}\ }\textbf {\bibinfo
  {volume} {86}} (\bibinfo {year} {2015})}\BibitemShut {NoStop}%
\bibitem [{\citenamefont {{Toptica Photonics AG}}(2017)}]{DL110}%
  \BibitemOpen
  \bibfield  {author} {\bibinfo {author} {\bibnamefont {{Toptica Photonics
  AG}}},\ }\href@noop {} {\enquote {\bibinfo {title} {{DigiLock 110: Digital
  Laser Locking}},}\ } (\bibinfo {year} {2017}),\ \bibinfo {note}
  {\url{http://www.toptica.com/products/tunable-diode-lasers/laser-locking-electronics/digilock-110-digital-locking/}}\BibitemShut
  {NoStop}%
\bibitem [{\citenamefont {Abitan}\ and\ \citenamefont
  {Skettrup.}(2005)}]{AS2005}%
  \BibitemOpen
  \bibfield  {author} {\bibinfo {author} {\bibfnamefont {H.}~\bibnamefont
  {Abitan}}\ and\ \bibinfo {author} {\bibfnamefont {T.}~\bibnamefont
  {Skettrup.}},\ }\href@noop {} {\bibfield  {journal} {\bibinfo  {journal} {{
  J. Opt. Pure Appl. Opt}}\ }\textbf {\bibinfo {volume} {7}},\ \bibinfo {pages}
  {7} (\bibinfo {year} {2005})}\BibitemShut {NoStop}%
\bibitem [{Red(2016)}]{RedPit}%
  \BibitemOpen
  \href@noop {} {\enquote {\bibinfo {title} {{Red Pitaya Project Website}},}\ }
  (\bibinfo {year} {2016}),\ \bibinfo {note}
  {\url{http://redpitaya.com/}}\BibitemShut {NoStop}%
\bibitem [{\citenamefont {Fenske}(2015)}]{JF15}%
  \BibitemOpen
  \bibfield  {author} {\bibinfo {author} {\bibfnamefont {J.}~\bibnamefont
  {Fenske}},\ }\emph {\bibinfo {title} {{Implementierung eines digitalen
  PID-Reglers mit dem Entwicklungsboard Red Pitaya}}},\ \href@noop {} {\bibinfo
  {type} {Bechelor's thesis}},\ \bibinfo  {school} {Ostfalia Hochschule f{\"u}r
  angewandte Wissenschaften}, \bibinfo {address} {Germany} (\bibinfo {year}
  {2015})\BibitemShut {NoStop}%
\bibitem [{\citenamefont {Kato}(1986)}]{kato_second-harmonic_1986}%
  \BibitemOpen
  \bibfield  {author} {\bibinfo {author} {\bibfnamefont {K.}~\bibnamefont
  {Kato}},\ }\href
  {http://ieeexplore.ieee.org/stamp/stamp.jsp?tp=&arnumber=1073097&isnumber=23096}
  {\bibfield  {journal} {\bibinfo  {journal} {‎IEEE J. Quant. Electron.}\
  }\textbf {\bibinfo {volume} {22}},\ \bibinfo {pages} {1013} (\bibinfo {year}
  {1986})}\BibitemShut {NoStop}%
\bibitem [{\citenamefont {Nikogosyan}(2005)}]{nikogosyan_nonlinear_2005}%
  \BibitemOpen
  \bibfield  {author} {\bibinfo {author} {\bibfnamefont {D.~N.}\ \bibnamefont
  {Nikogosyan}},\ }\href@noop {} {\emph {\bibinfo {title} {Nonlinear {Optical}
  {Crystals}: {A} {Complete} {Survey}}}}\ (\bibinfo  {publisher} {Springer},\
  \bibinfo {year} {2005})\BibitemShut {NoStop}%
\bibitem [{\citenamefont {Armstrong}\ \emph {et~al.}(1962)\citenamefont
  {Armstrong}, \citenamefont {Bloembergen}, \citenamefont {Ducuing},\ and\
  \citenamefont {Pershan}}]{armstrong_interactions_1962}%
  \BibitemOpen
  \bibfield  {author} {\bibinfo {author} {\bibfnamefont {J.~A.}\ \bibnamefont
  {Armstrong}}, \bibinfo {author} {\bibfnamefont {N.}~\bibnamefont
  {Bloembergen}}, \bibinfo {author} {\bibfnamefont {J.}~\bibnamefont
  {Ducuing}}, \ and\ \bibinfo {author} {\bibfnamefont {P.~S.}\ \bibnamefont
  {Pershan}},\ }\href {\doibase 10.1103/PhysRev.127.1918} {\bibfield  {journal}
  {\bibinfo  {journal} {Phys. Rev.}\ }\textbf {\bibinfo {volume} {127}},\
  \bibinfo {pages} {1918} (\bibinfo {year} {1962})}\BibitemShut {NoStop}%
\bibitem [{\citenamefont {Ashkin}, \citenamefont {Boyd},\ and\ \citenamefont
  {Dziedzic}(1966)}]{Ashkin_Resonant_SHG_1966}%
  \BibitemOpen
  \bibfield  {author} {\bibinfo {author} {\bibfnamefont {A.}~\bibnamefont
  {Ashkin}}, \bibinfo {author} {\bibfnamefont {G.}~\bibnamefont {Boyd}}, \ and\
  \bibinfo {author} {\bibfnamefont {J.}~\bibnamefont {Dziedzic}},\ }\href
  {\doibase 10.1109/JQE.1966.1074007} {\bibfield  {journal} {\bibinfo
  {journal} {‎IEEE J. Quant. Electron.}\ }\textbf {\bibinfo {volume} {2}},\
  \bibinfo {pages} {109} (\bibinfo {year} {1966})}\BibitemShut {NoStop}%
\bibitem [{\citenamefont {Boyd}\ and\ \citenamefont
  {Kleinman}(1968)}]{boyd_parametric_1968}%
  \BibitemOpen
  \bibfield  {author} {\bibinfo {author} {\bibfnamefont {G.~D.}\ \bibnamefont
  {Boyd}}\ and\ \bibinfo {author} {\bibfnamefont {D.~A.}\ \bibnamefont
  {Kleinman}},\ }\href {\doibase 10.1063/1.1656831} {\bibfield  {journal}
  {\bibinfo  {journal} {Journal of Applied Physics}\ }\textbf {\bibinfo
  {volume} {39}},\ \bibinfo {pages} {3597} (\bibinfo {year}
  {1968})}\BibitemShut {NoStop}%
\bibitem [{\citenamefont {Jurdik}\ \emph {et~al.}(2002)\citenamefont {Jurdik},
  \citenamefont {Hohlfeld}, \citenamefont {van Etteger}, \citenamefont
  {Toonen}, \citenamefont {Meerts}, \citenamefont {van Kempen},\ and\
  \citenamefont {Rasing}}]{jurdik_performance_2002}%
  \BibitemOpen
  \bibfield  {author} {\bibinfo {author} {\bibfnamefont {E.}~\bibnamefont
  {Jurdik}}, \bibinfo {author} {\bibfnamefont {J.}~\bibnamefont {Hohlfeld}},
  \bibinfo {author} {\bibfnamefont {A.~F.}\ \bibnamefont {van Etteger}},
  \bibinfo {author} {\bibfnamefont {A.~J.}\ \bibnamefont {Toonen}}, \bibinfo
  {author} {\bibfnamefont {W.~L.}\ \bibnamefont {Meerts}}, \bibinfo {author}
  {\bibfnamefont {H.}~\bibnamefont {van Kempen}}, \ and\ \bibinfo {author}
  {\bibfnamefont {T.}~\bibnamefont {Rasing}},\ }\href {\doibase
  10.1364/JOSAB.19.001660} {\bibfield  {journal} {\bibinfo  {journal} {JOSA B
  B}\ }\textbf {\bibinfo {volume} {19}},\ \bibinfo {pages} {1660} (\bibinfo
  {year} {2002})}\BibitemShut {NoStop}%
\bibitem [{\citenamefont {Gouy}(1890)}]{Gouy_1890}%
  \BibitemOpen
  \bibfield  {author} {\bibinfo {author} {\bibfnamefont {L.~G.}\ \bibnamefont
  {Gouy}},\ }\href@noop {} {\bibfield  {journal} {\bibinfo  {journal} {{C. R.
  Acad. Sci.}}\ }\textbf {\bibinfo {volume} {110}},\ \bibinfo {pages} {1251}
  (\bibinfo {year} {1890})}\BibitemShut {NoStop}%
\bibitem [{\citenamefont {Freegarde}\ \emph {et~al.}(1997)\citenamefont
  {Freegarde}, \citenamefont {Coutts}, \citenamefont {Walz}, \citenamefont
  {Leibfried},\ and\ \citenamefont {H\"{a}nsch}}]{Freegarde:97}%
  \BibitemOpen
  \bibfield  {author} {\bibinfo {author} {\bibfnamefont {T.}~\bibnamefont
  {Freegarde}}, \bibinfo {author} {\bibfnamefont {J.}~\bibnamefont {Coutts}},
  \bibinfo {author} {\bibfnamefont {J.}~\bibnamefont {Walz}}, \bibinfo {author}
  {\bibfnamefont {D.}~\bibnamefont {Leibfried}}, \ and\ \bibinfo {author}
  {\bibfnamefont {T.~W.}\ \bibnamefont {H\"{a}nsch}},\ }\href {\doibase
  10.1364/JOSAB.14.002010} {\bibfield  {journal} {\bibinfo  {journal} {J. Opt.
  Soc. Am. B}\ }\textbf {\bibinfo {volume} {14}},\ \bibinfo {pages} {2010}
  (\bibinfo {year} {1997})}\BibitemShut {NoStop}%
\bibitem [{\citenamefont {Watanabe}\ \emph {et~al.}(1991)\citenamefont
  {Watanabe}, \citenamefont {Hayasaka}, \citenamefont {Imajo}, \citenamefont
  {Umezu},\ and\ \citenamefont {Urabe}}]{Wata1991}%
  \BibitemOpen
  \bibfield  {author} {\bibinfo {author} {\bibfnamefont {M.}~\bibnamefont
  {Watanabe}}, \bibinfo {author} {\bibfnamefont {K.}~\bibnamefont {Hayasaka}},
  \bibinfo {author} {\bibfnamefont {H.}~\bibnamefont {Imajo}}, \bibinfo
  {author} {\bibfnamefont {J.}~\bibnamefont {Umezu}}, \ and\ \bibinfo {author}
  {\bibfnamefont {S.}~\bibnamefont {Urabe}},\ }\href
  {url="https://doi.org/10.1007/BF00325475"} {\bibfield  {journal} {\bibinfo
  {journal} {{Appl. Phys. B.}}\ }\textbf {\bibinfo {volume} {55}},\ \bibinfo
  {pages} {11} (\bibinfo {year} {1991})}\BibitemShut {NoStop}%
\bibitem [{\citenamefont {Polzik}\ and\ \citenamefont
  {Kimble}(1991)}]{polzik_frequency_1991}%
  \BibitemOpen
  \bibfield  {author} {\bibinfo {author} {\bibfnamefont {E.~S.}\ \bibnamefont
  {Polzik}}\ and\ \bibinfo {author} {\bibfnamefont {H.~J.}\ \bibnamefont
  {Kimble}},\ }\href {\doibase 10.1364/OL.16.001400} {\bibfield  {journal}
  {\bibinfo  {journal} {Opt. Lett.}\ }\textbf {\bibinfo {volume} {16}},\
  \bibinfo {pages} {1400} (\bibinfo {year} {1991})}\BibitemShut {NoStop}%
\bibitem [{\citenamefont {H{\"a}nsch}\ and\ \citenamefont
  {Couillaud}(1980)}]{hansch_laser_1980}%
  \BibitemOpen
  \bibfield  {author} {\bibinfo {author} {\bibfnamefont {T.~W.}\ \bibnamefont
  {H{\"a}nsch}}\ and\ \bibinfo {author} {\bibfnamefont {B.}~\bibnamefont
  {Couillaud}},\ }\href
  {http://www.sciencedirect.com/science/article/pii/0030401880900693}
  {\bibfield  {journal} {\bibinfo  {journal} {Opt. Commun.}\ }\textbf {\bibinfo
  {volume} {35}},\ \bibinfo {pages} {441} (\bibinfo {year} {1980})}\BibitemShut
  {NoStop}%
\bibitem [{\citenamefont {Drever}\ \emph {et~al.}(1983)\citenamefont {Drever},
  \citenamefont {Hall}, \citenamefont {Kowalski}, \citenamefont {Hough},
  \citenamefont {Ford}, \citenamefont {Munley},\ and\ \citenamefont
  {Ward}}]{drever_laser_1983}%
  \BibitemOpen
  \bibfield  {author} {\bibinfo {author} {\bibfnamefont {R.~W.~P.}\
  \bibnamefont {Drever}}, \bibinfo {author} {\bibfnamefont {J.~L.}\
  \bibnamefont {Hall}}, \bibinfo {author} {\bibfnamefont {F.~V.}\ \bibnamefont
  {Kowalski}}, \bibinfo {author} {\bibfnamefont {J.}~\bibnamefont {Hough}},
  \bibinfo {author} {\bibfnamefont {G.~M.}\ \bibnamefont {Ford}}, \bibinfo
  {author} {\bibfnamefont {A.~J.}\ \bibnamefont {Munley}}, \ and\ \bibinfo
  {author} {\bibfnamefont {H.}~\bibnamefont {Ward}},\ }\href
  {https://doi.org/10.1007/BF00702605} {\bibfield  {journal} {\bibinfo
  {journal} {Appl. Phys. B.: Lasers and Optics}\ }\textbf {\bibinfo {volume}
  {31}},\ \bibinfo {pages} {97} (\bibinfo {year} {1983})}\BibitemShut {NoStop}%
\bibitem [{\citenamefont {{Thorlabs Inc.}}(2017)}]{thorfrequ}%
  \BibitemOpen
  \bibfield  {author} {\bibinfo {author} {\bibnamefont {{Thorlabs Inc.}}},\
  }\href@noop {} {} (\bibinfo {year} {2017}),\ \bibinfo {note} {{Co-Fired
  Piezoelectric Actuator datasheet{
  \url{https://www.thorlabs.de}}}}\BibitemShut {NoStop}%
\bibitem [{\citenamefont {Sandberg}(1993)}]{JCS1993}%
  \BibitemOpen
  \bibfield  {author} {\bibinfo {author} {\bibfnamefont {J.~C.}\ \bibnamefont
  {Sandberg}},\ }\emph {\bibinfo {title} {Research Towards Laser Spectroscopy
  of Trapped Atomic Hydrogen}},\ \href@noop {} {\bibinfo {type} {Ph. d.
  thesis}},\ \bibinfo  {school} {MIT} (\bibinfo {year} {1993})\BibitemShut
  {NoStop}%
\bibitem [{\citenamefont {Idel}(2016)}]{AI2016}%
  \BibitemOpen
  \bibfield  {author} {\bibinfo {author} {\bibfnamefont {A.}~\bibnamefont
  {Idel}},\ }\emph {\bibinfo {title} {{Ein kompaktes Lasersystem zum
  sympathetischen K{\"u}hlen einzelner (Anti-) Protonen durch $^9Be^+$
  Ionen}}},\ \href@noop {} {\bibinfo {type} {Master's thesis}},\ \bibinfo
  {school} {Leibniz Universit{\"a}t Hannover}, \bibinfo {address} {Germany}
  (\bibinfo {year} {2016})\BibitemShut {NoStop}%
\bibitem [{\citenamefont {{Red Pitaya team}}(2017)}]{rp}%
  \BibitemOpen
  \bibfield  {author} {\bibinfo {author} {\bibnamefont {{Red Pitaya team}}},\
  }\href@noop {} {\enquote {\bibinfo {title} {{Red Pitaya web page}},}\ }
  (\bibinfo {year} {2017}),\ \bibinfo {note}
  {{\url{http://redpitaya.com/}}}\BibitemShut {NoStop}%
\bibitem [{\citenamefont {Xillinx}(2016)}]{Zynq-data}%
  \BibitemOpen
  \bibfield  {author} {\bibinfo {author} {\bibnamefont {Xillinx}},\ }\href@noop
  {} {\enquote {\bibinfo {title} {{Zynq-7000 All Programmable SoC Overview}},}\
  } (\bibinfo {year} {2016}),\ \bibinfo {note}
  {{\url{http://www.xilinx.com/support/documentation/data_sheets/ds190-Zynq-7000-Overview.pdf}}}\BibitemShut
  {NoStop}%
\bibitem [{\citenamefont {et. al.}(2014)}]{Zynq}%
  \BibitemOpen
  \bibfield  {author} {\bibinfo {author} {\bibfnamefont {L.~H.~C.}\
  \bibnamefont {et. al.}},\ }\href@noop {} {\enquote {\bibinfo {title} {{The
  Zynq Book - Embedded Processing with the ARM® Cortex®-A9 on the Xilinx®
  Zynq®-7000 All Programmable SoC}},}\ } (\bibinfo {year} {2014}),\ \bibinfo
  {note}
  {{\url{http://www.analog.com/media/en/technical-documentation/data-sheets/AD9763_9765_9767.pdf}}}\BibitemShut
  {NoStop}%
\bibitem [{\citenamefont {Technology}(2011)}]{ADC}%
  \BibitemOpen
  \bibfield  {author} {\bibinfo {author} {\bibfnamefont {L.}~\bibnamefont
  {Technology}},\ }\href@noop {} {\enquote {\bibinfo {title} {{LTC2145-14 -
  14-Bit, 125Msps Low Power Dual ADCs}},}\ } (\bibinfo {year} {2011}),\
  \bibinfo {note}
  {{\url{http://cds.linear.com/docs/en/datasheet/21454314fa.pdf}}}\BibitemShut
  {NoStop}%
\bibitem [{\citenamefont {Devices}(2011)}]{DAC}%
  \BibitemOpen
  \bibfield  {author} {\bibinfo {author} {\bibfnamefont {A.}~\bibnamefont
  {Devices}},\ }\href@noop {} {\enquote {\bibinfo {title} {Ad9767 - 14-bit, 125
  msps dual txdac+ digital-to-analog converters},}\ } (\bibinfo {year}
  {2011}),\ \bibinfo {note}
  {\url{http://www.analog.com/media/en/technical-documentation/data-sheets/AD9763_9765_9767.pdf}}\BibitemShut
  {NoStop}%
\bibitem [{\citenamefont {{Red Pitaya
  team}}(2016{\natexlab{a}})}]{kick_starter}%
  \BibitemOpen
  \bibfield  {author} {\bibinfo {author} {\bibnamefont {{Red Pitaya team}}},\
  }\href@noop {} {\enquote {\bibinfo {title} {{Red Pitaya: Open instruments for
  everyone}},}\ } (\bibinfo {year} {2016}{\natexlab{a}}),\ \bibinfo {note}
  {{\url{https://www.kickstarter.com/projects/652945597/red-pitaya-open-instruments-for-everyone/description}}}\BibitemShut
  {NoStop}%
\bibitem [{\citenamefont {{User LNEUHAUS}}(2016)}]{LNEUHAUS}%
  \BibitemOpen
  \bibfield  {author} {\bibinfo {author} {\bibnamefont {{User LNEUHAUS}}},\
  }\href@noop {} {\enquote {\bibinfo {title} {{Red Pitaya DAC performance}},}\
  } (\bibinfo {year} {2016}),\ \bibinfo {note}
  {{\url{https://ln1985blog.wordpress.com/2016/02/07/red-pitaya-dac-performance/}}}\BibitemShut
  {NoStop}%
\bibitem [{\citenamefont {{Red Pitaya team}}(2016{\natexlab{b}})}]{rpwiki_pid}%
  \BibitemOpen
  \bibfield  {author} {\bibinfo {author} {\bibnamefont {{Red Pitaya team}}},\
  }\href@noop {} {\enquote {\bibinfo {title} {{PID Controller}},}\ } (\bibinfo
  {year} {2016}{\natexlab{b}}),\ \bibinfo {note}
  {{\url{http://wiki.redpitaya.com/index.php?title=PID_controller}}}\BibitemShut
  {NoStop}%
\bibitem [{\citenamefont {{Red Pitaya team}}(2016{\natexlab{c}})}]{rp_mmap}%
  \BibitemOpen
  \bibfield  {author} {\bibinfo {author} {\bibnamefont {{Red Pitaya team}}},\
  }\href@noop {} {\enquote {\bibinfo {title} {{RedPitaya - FPGA memory map}},}\
  } (\bibinfo {year} {2016}{\natexlab{c}}),\ \bibinfo {note}
  {{\url{http://wiki.redpitaya.com/tmp/RedPitaya_HDL_memory_map.pdf}}}\BibitemShut
  {NoStop}%
\bibitem [{git(2016)}]{github}%
  \BibitemOpen
  \href@noop {} {} (\bibinfo {year} {2016}),\ \bibinfo {note} {{Sourcecode:
  \url{https://github.com/Julia-F/RedPitaya}}}\BibitemShut {NoStop}%
\bibitem [{\citenamefont {Jitschin}\ and\ \citenamefont
  {Meisel}(1979)}]{Jitschin_fast_1979}%
  \BibitemOpen
  \bibfield  {author} {\bibinfo {author} {\bibfnamefont {W.}~\bibnamefont
  {Jitschin}}\ and\ \bibinfo {author} {\bibfnamefont {G.}~\bibnamefont
  {Meisel}},\ }\href {\doibase 10.1007/BF00932394} {\bibfield  {journal}
  {\bibinfo  {journal} {Applied physics}\ }\textbf {\bibinfo {volume} {19}},\
  \bibinfo {pages} {181} (\bibinfo {year} {1979})}\BibitemShut {NoStop}%
\bibitem [{\citenamefont {Briles}\ \emph {et~al.}(2010)\citenamefont {Briles},
  \citenamefont {Yost}, \citenamefont {Cingöz}, \citenamefont {Ye},\ and\
  \citenamefont {Schibli}}]{briles_simple_2010}%
  \BibitemOpen
  \bibfield  {author} {\bibinfo {author} {\bibfnamefont {T.~C.}\ \bibnamefont
  {Briles}}, \bibinfo {author} {\bibfnamefont {D.~C.}\ \bibnamefont {Yost}},
  \bibinfo {author} {\bibfnamefont {A.}~\bibnamefont {Cingöz}}, \bibinfo
  {author} {\bibfnamefont {J.}~\bibnamefont {Ye}}, \ and\ \bibinfo {author}
  {\bibfnamefont {T.~R.}\ \bibnamefont {Schibli}},\ }\href {\doibase
  10.1364/OE.18.009739} {\bibfield  {journal} {\bibinfo  {journal} {Opt.
  Express}\ }\textbf {\bibinfo {volume} {18}},\ \bibinfo {pages} {9739}
  (\bibinfo {year} {2010})}\BibitemShut {NoStop}%
\bibitem [{\citenamefont {Goldovsky}, \citenamefont {Jouravsky},\ and\
  \citenamefont {Pe’er}(2016)}]{goldovsky_simple_2016}%
  \BibitemOpen
  \bibfield  {author} {\bibinfo {author} {\bibfnamefont {D.}~\bibnamefont
  {Goldovsky}}, \bibinfo {author} {\bibfnamefont {V.}~\bibnamefont
  {Jouravsky}}, \ and\ \bibinfo {author} {\bibfnamefont {A.}~\bibnamefont
  {Pe’er}},\ }\href {\doibase 10.1364/OE.24.028239} {\bibfield  {journal}
  {\bibinfo  {journal} {Opt. Express}\ }\textbf {\bibinfo {volume} {24}},\
  \bibinfo {pages} {28239} (\bibinfo {year} {2016})}\BibitemShut {NoStop}%
\bibitem [{MCA(2017)}]{MCAC}%
  \BibitemOpen
  \href@noop {} {\enquote {\bibinfo {title} {{Multi-component acceleration
  exciter PTB Working Group 1.71}},}\ } (\bibinfo {year} {2017}),\ \bibinfo
  {note}
  {\url{https://www.ptb.de/cms/en/ptb/fachabteilungen/abt1/fb-17/ag-171/research-and-development-01/multi-component-acceleration-exciter.html}}\BibitemShut
  {NoStop}%
\bibitem [{\citenamefont {{Red Pitaya team}}(2016{\natexlab{d}})}]{rpwiki_h}%
  \BibitemOpen
  \bibfield  {author} {\bibinfo {author} {\bibnamefont {{Red Pitaya team}}},\
  }\href@noop {} {\enquote {\bibinfo {title} {{Hardware Overview}},}\ }
  (\bibinfo {year} {2016}{\natexlab{d}}),\ \bibinfo {note}
  {{\url{http://wiki.redpitaya.com/index.php?title=Hardware_Overview}}}\BibitemShut
  {NoStop}%
\end{thebibliography}
%

\end{document}